\documentclass[a4paper,fleqn,usenatbib]{mnras}

\usepackage[T1]{fontenc}
\usepackage{ae,aecompl}

\usepackage{graphicx}	
\usepackage{amsmath}	
\usepackage{amssymb}	
\usepackage{natbib}
\usepackage{float}
\usepackage[usenames,dvipsnames]{xcolor}
\usepackage[normalem]{ulem}
\usepackage{multirow}

\newcommand{\mvir}{M_\mathrm{v}}
\newcommand{\rvir}{R_\mathrm{v}}

\newcommand{\vvir}{V_\mathrm{v}}
\newcommand{\vmax}{V_\mathrm{max}}

\newcommand{\msun}{\textnormal{M}_\odot}
\newcommand{\Msun}{\,\textnormal{M}_\odot}
\newcommand{\mstar}{M_\star}
\newcommand{\mhalo}{M_\mathrm{halo}}
\newcommand{\mpeak}{M_\mathrm{peak}}

\newcommand{\mpc}{\mathrm{Mpc}}
\newcommand{\Mpc}{\,\textnormal{Mpc}}
\newcommand{\kpc}{\mathrm{kpc}}
\newcommand{\pc}{\mathrm{pc}}

\newcommand{\hmpc}{h^{-1}\,\mathrm{Mpc}}

\newcommand{\kms}{{\rm km} \ {\rm s}^{-1}}
\newcommand{\lcdm}{$\Lambda$CDM}

\newcommand{\Rd}{R_\mathrm{d}}
\newcommand{\zd}{z_\mathrm{d}}
\newcommand{\Md}{M_\mathrm{d}}
\newcommand{\vrad}{V_\mathrm{rad}}
\newcommand{\vtan}{V_\mathrm{tan}}
\newcommand{\fb}{f_\mathrm{b}}
\newcommand{\dperi}{d_\mathrm{peri}}

\newcommand{\mf}{\texttt{m12f}}
\newcommand{\mi}{\texttt{m12i}}

\defcitealias{Sawala2016b}{S17}

\title[Not so lumpy after all]{Not so lumpy after all: modeling the depletion of dark matter subhalos by Milky Way-like galaxies}
\author[S. Garrison-Kimmel et al.]{Shea Garrison-Kimmel$^{1}$\thanks{$\!$sheagk@caltech.edu},
  Andrew Wetzel$^{1,2,3}$\thanks{Caltech-Carnegie Fellow},
  James S. Bullock$^{4}$, \and
  Philip F. Hopkins$^{1}$,
  Michael Boylan-Kolchin$^{5}$,
  Claude-Andr{\'e} Faucher-Gigu{\`e}re$^{6}$, \and
  Du{\v s}an Kere{\v s}$^{7}$,
  Eliot~Quataert$^{8}$,
  Robyn E. Sanderson$^{1}$,
  Andrew S. Graus$^{4}$, \and
  Tyler Kelley$^{4}$
\\
$\!\!$ $^1$TAPIR, California Institute of Technology, Pasadena, CA 91125, USA \\
$\!\!$ $^2$The Observatories of the Carnegie Institution for Science, Pasadena, CA 91125, USA \\
$\!\!$ $^3$Department of Physics, University of California, Davis, CA 95616, USA \\
$\!\!$ $^4$Center for Cosmology, Department of Physics and Astronomy, University of California, Irvine, CA 92697, USA \\
$\!\!$ $^5$Department of Astronomy, The University of Texas at Austin, 2515 Speedway, Stop C1400, Austin, TX 78712 \\
$\!\!$ $^6$Center for Interdisciplinary Exploration and Research in Astrophysics (CIERA) and Department of Physics and Astronomy, \\ Northwestern University, 2145 Sheridan Road, Evanston, IL 60208, USA \\
$\!\!$ $^7$Department of Physics, Center for Astrophysics and Space Sciences, University of California, San Diego, La Jolla, CA 92093, USA \\
$\!\!$ $^8$Department of Astronomy and Theoretical Astrophysics Center, University of California, Berkeley, CA 94720-3411, USA \\
}

\date{Accepted XXX. Received YYY; in original form ZZZ}

\pubyear{2017}

\begin{document}
\label{firstpage}
\pagerange{\pageref{firstpage}--\pageref{lastpage}}

\maketitle

\begin{abstract}
Among the most important goals in cosmology is detecting and quantifying small
($\mhalo\simeq10^{6-9}\Msun$) dark matter (DM) subhalos. Current probes around
the Milky Way (MW) are most sensitive to such substructure within $\sim20$~kpc
of the halo center, where the galaxy contributes significantly to the potential.
We explore the effects of baryons on subhalo populations in $\Lambda$CDM using
cosmological zoom-in baryonic simulations of MW-mass halos from the Latte simulation
suite, part of the Feedback In Realistic Environments (FIRE) project.  Specifically,
we compare simulations of the same two halos run using (1) DM-only (DMO), (2) full
baryonic physics, and (3) DM with an embedded disk potential grown to match the FIRE
simulation. Relative to baryonic simulations, DMO simulations contain $\sim2\times$
as many subhalos within 100 kpc of the halo center; this excess is $\gtrsim5\times$ within
25 kpc. At $z=0$, the baryonic simulations are completely devoid of subhalos down to
$3\times10^6\Msun$ within $15$ kpc of the MW-mass galaxy, and fewer than 20
surviving subhalos have orbital pericenters $<20~\kpc$.  Despite the complexities of 
baryonic physics, the simple addition of an embedded central disk potential to DMO
simulations reproduces this subhalo depletion, including trends with radius, remarkably
well. Thus, the additional tidal field from the central galaxy is the primary cause of
subhalo depletion. Subhalos on radial orbits that pass close to the central galaxy are
preferentially destroyed, causing the surviving population to have tangentially
biased orbits compared to DMO predictions. Our method of embedding a potential in
DMO simulations provides a fast and accurate alternative to full baryonic simulations,
thus enabling suites of cosmological simulations that can provide accurate and statistical
predictions of substructure populations.
\end{abstract}




\section{Introduction}
\label{sec:intro}

One of the strongest predictions of \lcdm\ (cosmological constant with cold dark matter) is
that dark matter clusters hierarchically: large halos that host Milky Way (MW)-size galaxies
are filled with smaller, self-bound clumps known as subhalos. The highest-resolution cosmological
simulations of MW-size halos in the \lcdm\ paradigm have demonstrated that dark matter (DM)
clumps exist at all resolved masses \citep[e.g.][]{Aquarius,VL2,GHALO,ELVIS,Griffen2016}.

While at least some of these subhalos are presumed to host faint satellite galaxies,
the `missing satellites' problem \citep{Klypin1999MissingSats,Moore1999} points out a
sharp discrepancy between the flat luminosity function of observed satellites and the steep,
ever-rising mass function of subhalos predicted from numerical simulations. Though the
discrepancy can be largely eliminated by invoking gas heating from reionization suppressing
star formation in the early Universe \citep{Bullock2000,Somerville2002} and observational
incompleteness \citep{Tollerud2008}, this solution demands that subhalos with bound
masses smaller than $\sim10^8 \Msun$ should be \emph{dark} and that thousands of
$\sim10^{6} \Msun$ subhalos should be entirely devoid of stars.

Confirming the existence of these tiny, dark subhalos, and further determining
their mass function, would simultaneously provide an astounding confirmation of
the \lcdm\ theory and rule out large classes of warm dark matter models and
inflationary models that predict a cut-off in the power spectrum at low masses
\citep{KL2000,Bode2001,ZB03,Horiuchi2016,Bozek2016,Bose2016}. Because these subhalos are
dark, however, they must be identified indirectly. Around the MW, the best possibilities
for detecting dark substructure are through perturbations to the stellar disk 
\citep[e.g.][and references therein]{Quinn1993,Feldmann2015} and via gaps or kinematic 
distortions in dynamically cold stellar streams formed from disrupting globular clusters, 
such as Palomar-5 and GD-1 \citep[e.g.][and references therein]{Johnston2002,Koposov2010,Carlberg2012,Ngan2015}. 
In fact, \citet{Bovy2016} recently claimed a measurement of $10^{+11}_{-6}$ subhalos of mass
$3\times10^6$--$10^9 \Msun$ within $20~\kpc$ of the MW via the Palomar-5 stream.
\citet{Amorisco2016}, however, argued that giant molecular clouds could also be responsible
for some of the fluctuations, and \citet{Ibata2016} reported a null detection using the
same stream. Currently undetected stellar streams may provide more information about dark
substructure around the MW \citep{Ngan2016}. Around larger, more distant galaxies, dark
substructures may be revealed by gravitational lensing anomalies from background sources
\citep{Mao98,dalal2002,Vegetti2010,MacLeod2013,Nierenberg2014,Hezaveh2016}, particularly with
the upcoming instruments on the \textit{James Webb Space Telescope} \citep[\textit{JWST};][]{MacLeod2013}. These lensing studies are
sensitive to substructure within a projected Einstein radius that is typically $\sim5$--$10~\kpc$
in size \citep{dalal2002,Fiacconi2016}.

Making predictions for these observations, and thus using them to constrain the
properties of DM on small scales, requires a \emph{statistical} sample of halos
simulated in \lcdm\ with sufficient resolution (particle mass $\lesssim 10^5 \Msun$)
to identify the tiny subhalos of interest. While several such simulations exist in
the literature \citep[e.g.][]{Aquarius,VL2,GHALO,Mao2015,Griffen2016}, the vast
majority are purely collisionless (DM-only; DMO), given their low computational
cost compared with fully baryonic simulations. However, DMO simulations are
problematic because they miss critical baryonic physics, including the simple presence
of a central galaxy in the halo, near the very regions where observational probes
of substructure are most sensitive.  While the smallest subhalos are likely devoid
of baryons, they nonetheless should be dynamically influenced by the central galaxy
\citep{Taylor2001,Hayashi2003,Berezinsky2006,Read2006a,Read2006b,Penarrubia2010,DOnghia2010}.
In fact, the mere existence of a stellar halo around the MW implies a significant
population of destroyed dwarf galaxies.

Furthermore, observational estimates for the masses of larger subhalos, which host
luminous dwarf galaxies and therefore can be identified through direct observations,
are well below expectations from DMO simulations of MW-like halos.  The magnitude
of this discrepancy, which is known as the `too-big-to-fail' problem (\citealp{BK2011,BK2012}; 
see also \citealp{Tollerud2014} for similar results for M31 and \citealp{ELVISTBTF,Papastergis2015}
for field dwarf galaxies), depends on the adopted mass of the MW \citep{diCintio2011,Wang2012,Vera-Ciro2013},
but internal baryonic processes (such as rapid gaseous outflows driven by bursty stellar feedback
that can turn the cuspy DM profiles predicted by collisionless simulations into cored profiles with 
central masses at $r \lesssim 500~\pc$ that are consistent with the observations) appear important
for resolving TBTF \citep[e.g.][]{Navarro1996,Read2005,Mashchenko2008,Pontzen2012,Governato2012,Amorisco2013feedback,Gritschneder2013,Onorbe2015,Chan2015}.
As \citet{BrooksZolotov2012} showed, the increased tidal forces from the central
galaxy also can reduce the central masses of such dwarf galaxies even further,
potentially eliminating the disagreement between theory and observations entirely
\citep[also see][]{Read2006a,Read2006b}. Additionally, the central galaxy potential
may completely destroy some of the large, dense subhalos, effectively lowering
theoretical predictions for the central masses of the halos expected to host the
luminous dwarfs by placing them in correspondingly lower mass, and thus more
abundant, subhalos \citep[also see][]{GK2016}. However, the amount of destruction that
can be unambiguously attributed to the central galaxy remains uncertain.

Recently, the `Latte' simulation \citep{Wetzel2016}, part of the Feedback In Realistic
Environments (FIRE) project, achieved baryonic mass resolution of $\sim7000 \Msun$ for
a MW-mass halo run to $z = 0$.
The Latte simulation resolves subhalos down to mass $\sim10^{6} \Msun$, and as
\citet{Wetzel2016} showed, the initial Latte simulation produces a population of satellite
dwarf galaxies that agrees with a wide variety of observations around the MW and M31: the
distributions of stellar masses, velocity dispersions, and star formation histories, and the
relationship between stellar mass and metallicity all agree well those of the MW satellites.
Thus, it does not suffer from either the missing satellites or too-big-to-fail problems, at
least for resolved satellite dwarf galaxies ($\mstar \gtrsim 10^5 \Msun$).  Using the APOSTLE
simulations of Local Group-like MW-M31 pairs, \citet{Sawala2016b} also recently found good
observational agreement for satellite stellar masses and subhalo circular velocities using
different treatments of baryonic physics.  While these results from baryonic simulations are
quite promising, such simulations are sufficiently expensive to prohibit large parameter surveys
and statistical samples.

In this work, we extend the initial analysis of \citet{Wetzel2016} to study substructure
populations down to the smallest mass scales of relevance for current dark substructure
searches ($\sim10^{6} \Msun$).  We also present a second simulation of a MW-mass galaxy
in the Latte suite.  We will show that properly accounting for the effects of baryons is
essential for accurately predicting subhalo populations, even for completely dark subhalos
that have no stars. Motivated by previous work that suggested that adding an analytic potential
or other inexpensive modification(s) to DMO simulations could capture the key baryonic effects
on dark subhalos \citep[e.g.][]{DOnghia2010,Zhu2016,Sawala2016b,Errani2016,Jethwa2016}, we apply
a method for inserting an approximate analytic description of the gravitational potential of
the galaxy that forms at the center of each MW-mass halo into a cosmological DMO simulation.

Our approach is particularly interesting for testing the underlying physical drivers, because
we both calibrate our input central disk model and benchmark our results against our fully
baryonic simulations, which reproduce many observable properties of the satellite populations
around the MW and M31. As we will show, many key differences in subhalos populations between
DMO and baryonic simulations can be unambiguously attributed to the presence of the central
galaxy potential. This success of embedding a central galaxy potential also demonstrates that
the substructure populations predicted by cosmological DMO simulations can be significantly
improved (i.e. brought into better agreement with fully baryonic simulations) at minimal 
CPU cost.

In \S~\ref{sec:sims}, we discuss the simulations and detail our method of inserting an
embedded potential into the center of the host; \S\ref{sec:results} explores subhalo
population statistics with and without a forming galaxy and presents trends with radius
in subhalo depletion.  We discuss further implications of our results in \S\ref{sec:discussion}
and conclude in \S\ref{sec:conclusions}.

Throughout this work, we use $h = 0.702$, $\Omega_\mathrm{m} = 0.272$, $\Omega_\mathrm{b} = 0.0455$,
and $\Omega_\Lambda = 0.728$.

\section{Simulations}
\label{sec:sims}

\begin{figure*}
\centering
\setlength{\tabcolsep}{1pt}
\renewcommand{\arraystretch}{0.75}
\begin{tabular}{ccc}
\includegraphics[width=0.325\textwidth]{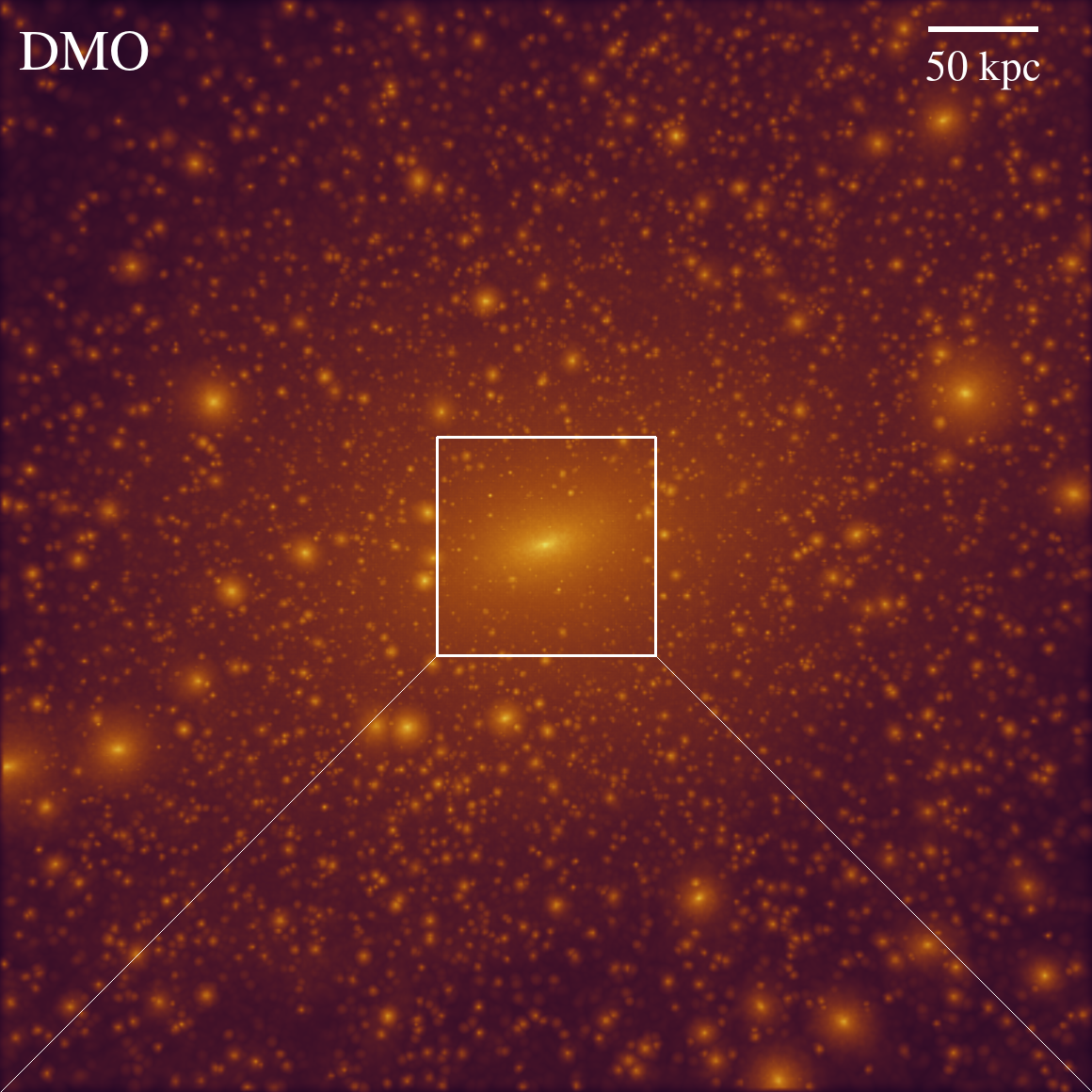} &
\includegraphics[width=0.325\textwidth]{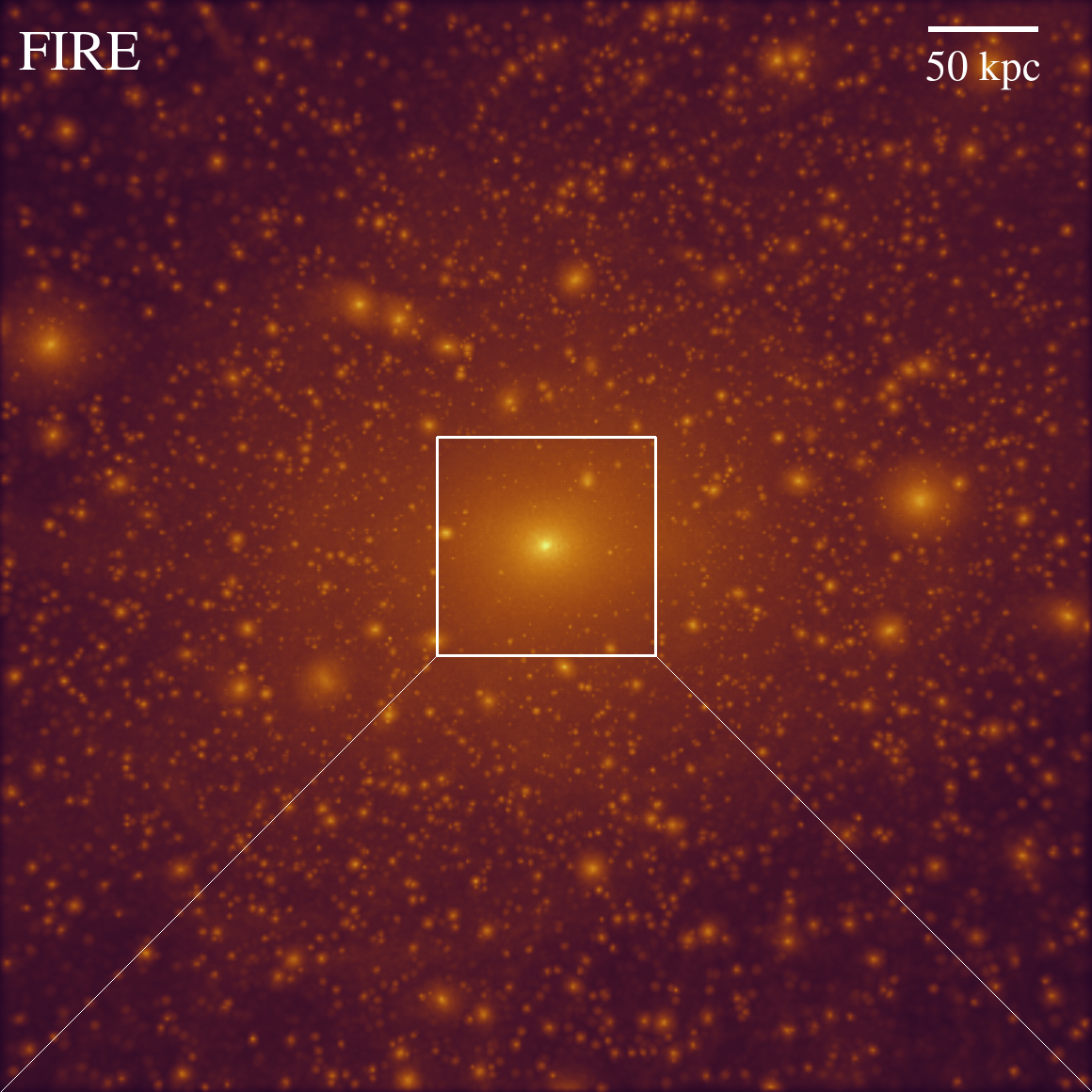} &
\includegraphics[width=0.325\textwidth]{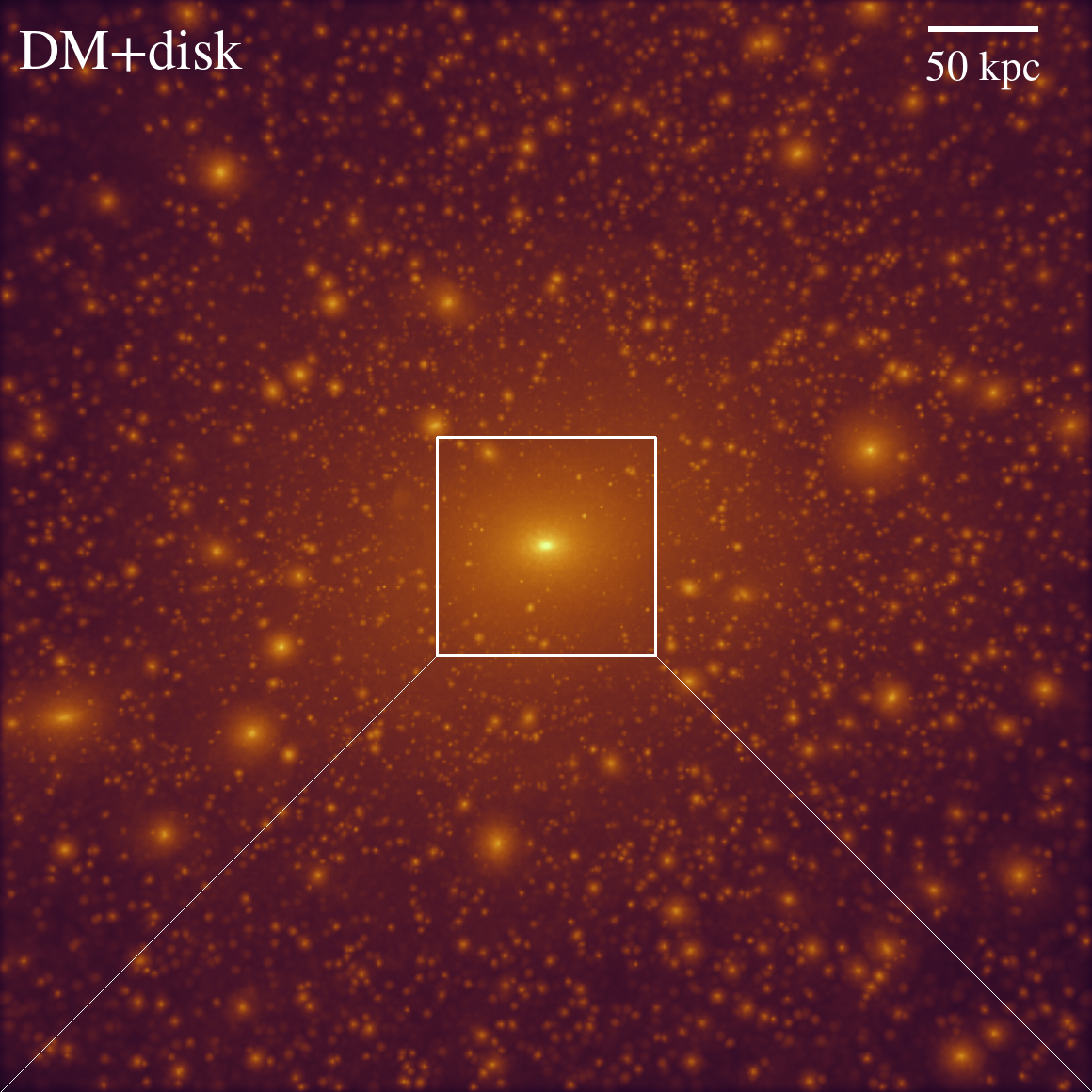} \\
\includegraphics[width=0.325\textwidth]{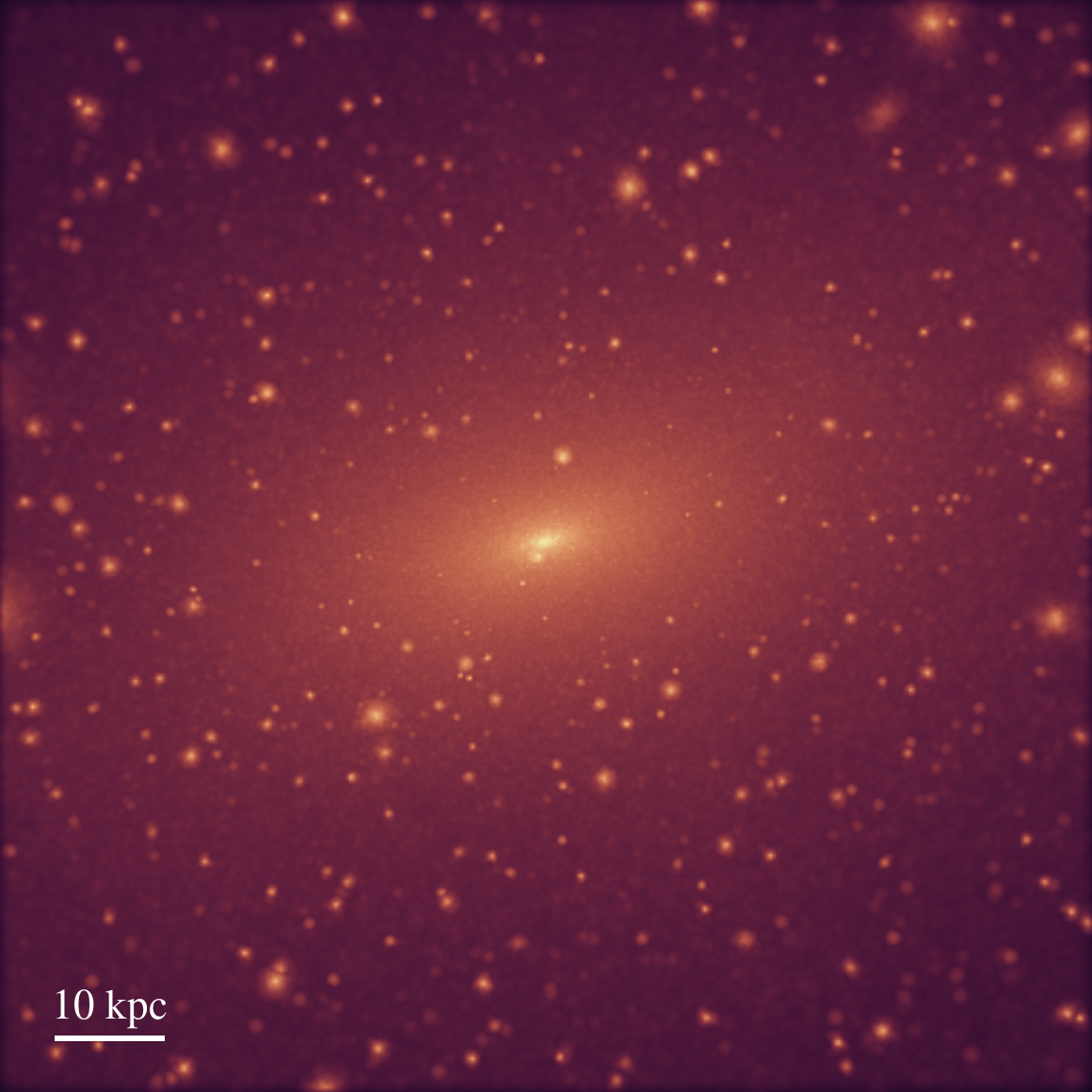} &
\includegraphics[width=0.325\textwidth]{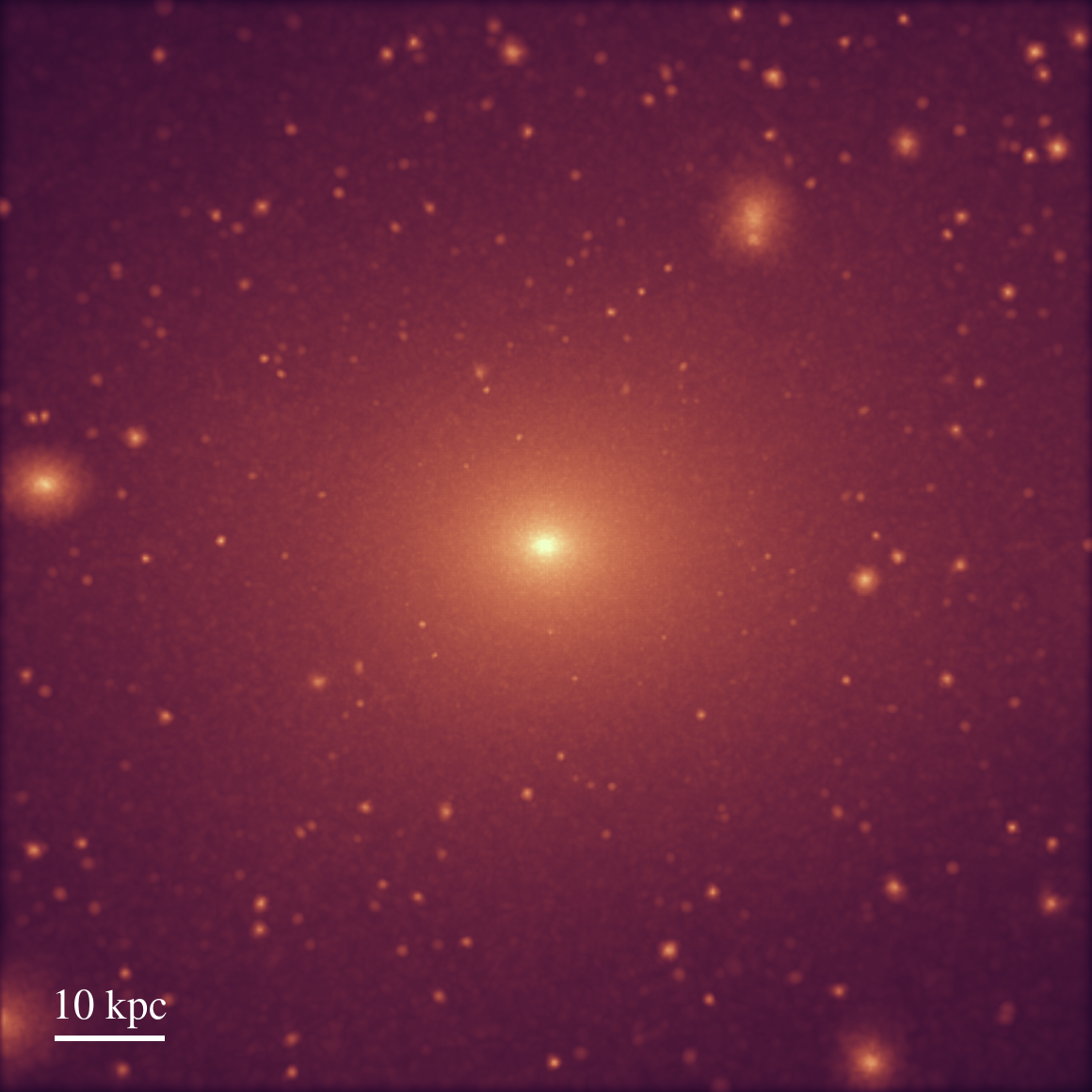} &
\includegraphics[width=0.325\textwidth]{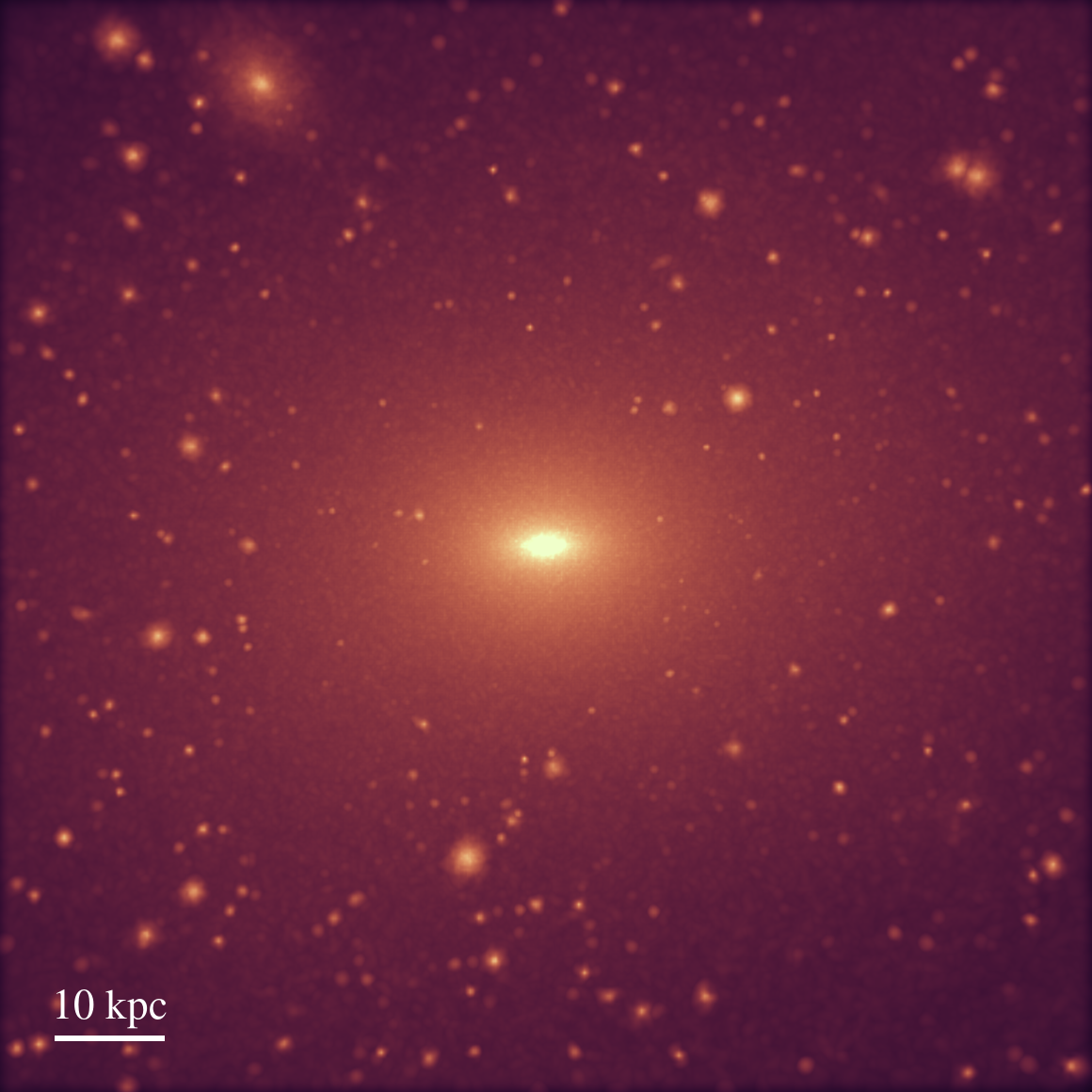}
\end{tabular}
\caption{
Visualizations of dark matter (DM) in the Latte \texttt{m12i} halo. Coloring indicates $\log_{10}$
of the local dark matter density. From left to right, the columns show the dark matter-only (DMO)
simulation, the fully baryonic simulation using FIRE physics, and the dark matter-only run that
adds an analytic, embedded disk potential to the halo center (DM+disk), where the disk properties
are matched to the baryonic simulation. The \textit{top} row illustrates a cube $500~\kpc$ on a
side, while the \textit{bottom} row zooms in on a cube $100~\kpc$ across. The presence of the central
galaxy (either real or embedded) leads to an enhancement in the DM density at the center. Substructure
counts are roughly similar on large scales in all cases (top row), but the tidal field of the central
galaxy eliminates many subhalos within $\sim50~\kpc$ (bottom row). Although the embedded disk potential
does not capture all of the effects of baryons, it does effectively capture subhalo depletion in the inner
halo, where searches for dark substructure via lensing or stellar streams are most sensitive. We
quantify these differences in Figures~\ref{fig:vmaxdist}~--~\ref{fig:massfuncs}.}
\label{fig:vizes}
\end{figure*}

All of our simulations are cosmological and employ the `zoom-in' technique \citep{Katz1993,Onorbe2014}.
We run all of our simulations in the same cosmological volume as the AGORA project
\citep{Kim2014}, with box length of $60 \hmpc = 85.5~\mpc$.  We choose each high-resolution
region to contain a single MW-mass ($\mhalo \sim10^{12} \Msun$) halo at $z = 0$ that has no
neighboring halos of similar or greater mass within $3 \Mpc$. We focus on two such halos,
designated as \mi\ and \mf, which are part of the Latte sample from the FIRE project.  \mi\
was presented in \citet{Wetzel2016}; \mf\ was simulated with identical parameters using the
same pipeline and will be described in detail in Wetzel et al. (in preparation).  We chose
both halos based only on their virial mass (see Table~\ref{tab:sims}) and not based on their
formation/merger history or subhalo population.

We ran all simulations using \texttt{GIZMO}
\citep{GIZMO}\footnote{\url{http://www.tapir.caltech.edu/~phopkins/Site/GIZMO.html}},
which uses an updated version of the TREE+PM gravity solver included in
\texttt{GADGET-3} \citep{Springel2005}.  We identify halo centers and create halo 
catalogs with \texttt{AHF} \citep{AHF} and build merger trees using\texttt{consistent-trees} 
\citep{ctrees}.  We generated initial conditions for the DMO and baryonic simulations at $z = 100$ 
using \texttt{MUSIC} \citep{MUSIC} with second-order Lagrangian perturbation theory.  For the
embedded disk simulations, we generated initial conditions using the snapshot at $z = 3$
from the DMO simulations.

The properties of the two host halos, along with the number of resolved subhalos
identified by \texttt{AHF} within several radial cuts of that host, are listed in
Table~\ref{tab:sims}. The first column lists the names of the simulations: those
identified with `\texttt{dmo}' are purely collisionless, while the simulations
labeled as `\texttt{disk}' are DMO with an embedded galactic potential; the other
rows show the FIRE baryonic simulations. The remaining columns list the \citet{Bryan1998}
virial masses, maximum circular velocities, $\vmax$, and virial velocities
($\sqrt{G\mvir/\rvir}$) of the hosts, here calculated from the full particle distribution
(including gas and stars in the baryonic simulations), the total number of subhalos with
$\vmax > 5\kms$ that survive to $z=0$ within $25$, $50$, $100$, and $300~\kpc$ of the
halo center, and the fraction of the host mass that resides in self-bound subhalos.

\begin{table*}
\centering
\begin{tabular}{lcccccccc}
    Simulation   &   $\mvir$    &   $\vmax$    & $\vvir$    & $N_\mathrm{sub}$  &     $N_\mathrm{sub}$  & $N_\mathrm{sub}$ & $N_\mathrm{sub}$ & $f_\mathrm{bound}$ \\
                 &   ($10^{12} \Msun$)    & ($\mathrm{\kms}$)  & ($\mathrm{\kms}$) & $(<25\,\mathrm{kpc})$ & $(<50\,\mathrm{kpc})$  & $(<100\,\mathrm{kpc})$ & $(<300\,\mathrm{kpc})$ & \\
     \hline\hline

     \mf               & 1.6 & 183 & 149 & 4 & 62 & 266 & 1482   &  0.037 \\
     \mf\texttt{-dmo}  & 1.6 & 177 & 150 & 42 & 204 & 654 & 2423 & 0.081 \\
     \mf\texttt{-disk} & 1.7 & 195 & 153 & 13 & 97 & 379 & 1854  & 0.049 \\
     \hline
     \mi              & 1.1 & 163 & 134 & 9 & 57 & 370 & 1432    & 0.056 \\
     \mi\texttt{-dmo} & 1.1 & 162 & 134 & 39 & 200 & 671 & 2069  & 0.078 \\
     \mi\texttt{-disk} & 1.2 & 191 & 136 & 6 & 108 & 473 & 1712  & 0.062 \\
     \hline
     \mi\texttt{-disk-}2$\Md$ & 1.2 & 220 & 138 & 5 & 96 & 455 & 1619  & 0.051 \\
     \mi\texttt{-disk-}$M_\star + M_\mathrm{gas}$ & 1.2 & 188 & 136 & 9 & 45  & 352 & 2142  & 0.062 \\
     \mi\texttt{-disk-}2$\Rd$ & 1.2 & 189 & 136 & 16 & 129 & 485 & 1738  & 0.062 \\
     \mi\texttt{-disk-}$\zd$0 & 1.2 & 183 & 136 & 19 & 120 & 529 & 1733  & 0.061 \\
     \mi\texttt{-Hernquist}   & 1.2 & 216 & 136 & 10 & 70  & 385 & 2214  & 0.066 \\
\end{tabular}
\caption{
Properties of the two host halos, \mi\ \citep[first presented in][]{Wetzel2016} and \mf\,
(first presented here and to be presented in more detail in Wetzel et al., in preparation).
`\texttt{-dmo}' refers to a dark matter-only (DMO) simulation of the same halo, while
`\texttt{-disk}' indicates a dark matter + embedded disk potential simulation; the
entries without an appendix represent the FIRE baryonic simulations. The `2$\Md$,'
`$M_\star + M_\mathrm{gas}$,' `2$\Rd$,' `$\zd0$,' and `\texttt{Hernquist}' simulations 
below the break indicate runs discussed in \S\ref{ssec:params}, which include factors of 
$\lesssim2$ changes  to the disk parameters, or, in the final case, a spherically symmetric 
potential with the same mass. The columns indicate the name of the simulation, the mass of 
the host halo \citep[using the virial definition of][]{Bryan1998}, the maximum circular velocity, 
$\vmax$, of the host halo (including only DM), the virial velocity, $\vvir$, of the host halo 
(defined as $\sqrt{G\mvir/\rvir}$, where $\rvir$ is the virial radius), the number of subhalos 
within $25$, $50$, $100$, and $300~\kpc$ at $z=0$ with $\vmax > 5~\kms$ (corresponding roughly to 
$M_{\rm bound} \simeq 5 \times 10^6 \Msun$), and $f_\mathrm{bound}$, the total mass in resolved 
subhalos divided by the virial mass of the host.  $\mvir$ and $\vvir$ for the baryonic simulations 
include the contributions of gas and stars.  However, we list $\vmax$ based only on DM for these runs 
because the circular velocity curve peaks at small radii ($r \simeq 1.6~\kpc$) when including baryons 
in the mass profile; in this case, \mi\ reaches $\vmax = 279~\kms$ while \mf\ reaches $\vmax = 283~\kms$.}
\label{tab:sims}
\end{table*}

\subsection{Baryonic simulations}
\label{ssec:hydrosims}

The baryonic simulation of \mi\ analyzed here is the same run presented in
\citet{Wetzel2016}; \mf\ was simulated with identical code and parameters.
The physics and numerical prescriptions are therefore given in \citet{Wetzel2016}.
Briefly, however, the baryonic simulations are part of the Feedback In Realistic Environments
\citep[FIRE;][]{FIRE} project\footnote{\url{http://fire.northwestern.edu}}.
Specifically, they use the updated FIRE-2 code, which features identical physics as FIRE-1
but incorporates several \textit{numerical} improvements.  In particular, FIRE-2 adopts the
new mesh-free finite mass (`MFM') method for more accurate hydrodynamics \citep{GIZMO}.
We model radiative heating and cooling from  $10 - 10^{10}\,$K
\citep[following \texttt{CLOUDY} tabulations;][]{Ferland1998},
and accounting for self-shielding and photo-heating both by a UV background
\citep[from][]{FaucherGiguere2009} and by local sources. Star formation occurs
in self-gravitating gas (according to the criterion in \citealp{Hopkins2013}) that
is also molecular and self-shielding \citep[following][]{Krumholz2011}, Jeans unstable,
and exceeds a minimum density threshold $n_{\rm sf} > 1000\,{\rm cm^{-3}}$.  A star
particle is then spawned probabilistically from a gas particle, inheriting the same
mass and metallicity. The simulations follow several stellar feedback mechanisms,
including (1) local and long-range momentum flux from radiation pressure
(in the initial UV/optical single-scattering, and from re-radiated light in the IR),
(2) energy, momentum, mass, and metal injection from core-collapse and Ia supernovae,
as well as stellar mass loss from OB and AGB stars, and (3) photo-ionization and
photo-electric heating. Every star particle is treated as a single stellar population
with a mass, age, and metallicity. We tabulate all feedback event rates, luminosities
and energies, mass-loss rates, and other quantities directly from stellar evolution
models \citep[\texttt{STARBURST99} v7.0;][]{Leitherer1999} assuming a \citet{Kroupa2001}
initial mass function (IMF).

Full details of FIRE-2 are provided in \citet{FIRE2}.  The source code and
numerical parameters of our baryonic simulations are {\em exactly} identical to those in all
FIRE-2 simulations \citep{Wetzel2016, Su2016, Fitts2016}.

The FIRE simulations have been shown to reproduce a wide variety of observables,
including the relationships between stellar mass and halo mass, the Kennicutt-Schmidt
law, bursty star formation histories, the star forming main sequence
\citep{FIRE}, galactic winds \citep{Muratov2015, Muratov2016}, the gas and stellar phase
$\mstar$-metallicity relations \citep{Ma2016a}, the $\mstar$-size relation
\citep{ElBadry2016}, the HI content of galaxy halos at both low and high redshift
\citep{FG2015, FG2016, Hafen2016}, and the structure and star formation histories of
isolated dwarf galaxies \citep{Onorbe2015,Chan2015,Fitts2016}.  Moreover, in simulations
of MW-mass halos, in addition to forming a realistic MW-like galaxy in terms of stellar
mass and disk morphology \citep{Wetzel2016, Ma2016b}, the FIRE model yields reasonable
populations of dwarf galaxies around those galaxies, in terms of the distributions of
stellar masses and velocity dispersions, as well as a wide range of star formation
histories that agree well with those of the actual MW satellites.

Both \mi\ and \mf\ form thin, radially extended stellar disks with
$\mstar(R<R_{90}, z<z_{90}) = 6.2 \times 10^{10} \Msun$ and $7.5 \times 10^{10} \Msun$,
respectively, where $R_{90}$ and $z_{90}$ are the radius and height that contain 90\% of the mass.
Thus, these galaxies are comparable to, if slightly more massive than, the MW in stars
\citep{BlandHawthorn2016}.  At $z = 0$, the total gas fraction, $M_{\rm gas} / (\mstar + M_{\rm gas})$,
within $R_{90}$ and $z_{90}$ is 13\% for \mi\ and 15\% for \mf.

The gravitational force softenings and kernel smoothing lengths for gas particles are fully
adaptive and conservative \citep[following][]{Price2007}. Hydrodynamic smoothings and gravitational
force softenings are always self-consistently matched. The minimum gas smoothing/Plummer equivalent
softening achieved in both simulations is $\epsilon_\mathrm{gas,\,min} = 1~\pc$ (corresponding to a
density of $n_{\rm gas} \approx \! 10^7\,\mathrm{cm}^{-3}$), thus ensuring that dense, star-forming regions are
well resolved. We choose softenings for the DM particles to be comparable to the typical gas softening
in the host galaxy's disk: $\epsilon_{\rm DM} = 20~\pc$.  The softenings for the stars are
$\epsilon_{\rm stars} = 8~\pc$, chosen to match the gas softening at the density threshold for
star-forming regions, $n_{\rm sf} > 1000~\mathrm{cm}^{-3}$.  All (minimum) softening lengths
quoted here are fixed in physical units after $z = 9$, and evolve comoving with the scale factor
prior to that redshift.  Each simulations is initialized at an `effective' resolution of
$2 \times 8192^3$ particles within the box, resulting in a dark matter particle mass of
$m_{\rm DM} = 3.5 \times 10^4 \Msun$ and a (initial) gas or star particle mass of
$m_{\rm gas} = 7.1 \times 10^{3} \Msun$.

\subsection{Dark matter-only simulations}
\label{ssec:DMOsims}

The dark matter-only (DMO) simulations are identical to the baryonic simulations,
except that they include only dark matter particles, and the baryonic mass is included
in the dark matter particles.  Consequently, the particle masses are larger by a factor
of $1/(1-\fb)$, where $\fb = \Omega_{b}/\Omega_\mathrm{m} \simeq 0.17$ is the cosmic
baryon fraction: $m_{\rm DM,\,DMO} = 4.2 \times 10^4 \Msun$.  This increased particle mass
has non-trivial effects on comparisons of DMO and baryonic simulations, both directly, as
in the case of subhalo mass functions, and indirectly, through implied resolution cuts.
This difference is particularly relevant for low-mass subhalos, which have lost essentially
all of their baryonic mass by $z = 0$ from cosmic reionization, feedback-driven
gas heating, and ram-pressure stripping within the host halo. Thus, a given subhalo
may contain an identical number of DM particles in the collisionless and baryonic
simulations, but be more massive in the former. Therefore, for low-mass subhalos, which are
expected to have lost (nearly) all of their baryonic mass, the most physical way to quote
their masses in DMO simulations is to correct for this (presumed) baryonic mass loss. We
thus multiply all particle and subhalo masses by $1-\fb$ in all post-processing (after halo
finding), and we similarly suppress the maximum circular velocities of all subhalos $\vmax$
by $\sqrt{1-\fb}$ \citep[similar to][]{Zolotov2012}. Finally, we perform halo finding only
on the DM particles in the baryonic simulations, to achieve the fairest comparison.
Therefore, the differences between baryonic and DMO simulations that we quote here
are somewhat smaller than quoted in \citet{Wetzel2016}, who did not include this
correction.

\subsection{Embedded disk potentials}
\label{ssec:embedding}

To include the effects of the disk of the central galaxy -- which grows naturally
within baryonic simulations -- in our DMO simulations, we add an additional
gravitational acceleration to every particle active during each timestep, as
given by a \citet{Miyamoto1975} potential:
\begin{equation}
    \Phi(R,z) = \frac{G\Md}{\left\{\Delta R^2 + \left[\Rd + \left(\Delta z^2 + \zd^2\right)^{1/2}\right]^2\right\}^{1/2}},
\label{eqn:phi}
\end{equation}
where $\Delta R$ and $\Delta z$ indicate the relative position from the center of the
potential ($x_{\rm disk}, y_{\rm disk}, z_{\rm disk}$) in cylindrical coordinates:
$\Delta R^2 = (x - x_\mathrm{disk})^2 + (y - y_\mathrm{disk})^2$;
$\Delta z^2 = (z - z_\mathrm{disk})^2$. This potential has three parameters: $\Rd(z)$,
the disk scale length, $\zd(z)$, the disk scale height, and $\Md(z)$, the total mass
in the potential.  Importantly, the acceleration from the Miyamoto-Nagai disk is
analytic: in the plane of the disk, taken here to be the $x-y$ plane, it is
\begin{equation}
    a_x = \frac{-G\Md}{\left\{\Delta R^2 + \left[\Rd + \left(\Delta z^2 + \zd^2\right)^{1/2}\right]^2\right\}^{3/2}} \Delta x,
\label{eqn:accelxy}
\end{equation}
where $\Delta x =  (x - x_\mathrm{disk})$, and similarly in the $y$ direction.
The acceleration along the minor axis, taken to be the $z$ direction, is
\begin{equation}
    a_z = \frac{-G\Md \left[\Rd+\left(\Delta z^2+\zd^2\right)^{1/2}\right]}{\left\{\Delta R^2 + \left[\Rd + \left(\Delta z^2 + \zd^2\right)^{1/2}\right]^2\right\}^{3/2}\left(\Delta z^2+\zd^2\right)^{1/2}} \Delta z.
\label{eqn:accelz}
\end{equation}

Within the DMO simulation, we track the center of the MW-mass host halo across cosmic time
using a single massive ($m_\mathrm{p} = 10^8 \Msun/h$) particle inserted at the center of
the main branch of the host at $z = 3$, using a large softening length
($\epsilon = 3.7$~physical~$\kpc$).  This effectively acts as a small bulge in the center
of the galaxy. Because this particle is significantly more massive than the high-resolution
particles that comprise the remainder of the halo, dynamical friction acts to keep it near the
center of the host.\footnote{At $z = 0$, the particle is within $12~\pc$ of the center of
\mi\ found by \texttt{AHF} and within $235~\pc$ of \mf, even without including the particle
in the halo finding.}  The position of this particle at each timestep determines the center
of the disk potential. To minimize computational cost, $(x,y,z)$ offsets from this particle
are computed for all other particles while traversing the gravity tree.

In our embedded disk simulations, we allow this disk potential to evolve over time to match
the stellar disk that forms in the corresponding baryonic simulation of the same system. We
first embed the disk potential at $z = 3$ ($\sim11.5~\mathrm{Gyr}$ ago), initializing with
parameters ($\Md$, $\Rd$, and $\zd$) obtained by jointly fitting the average density profile
of each baryonic simulation's stellar disk along the major and minor axes, defined by solving for
the eigenvectors of the moment of inertia tensor of all stars within 20 kpc. We then linearly
interpolate each parameter in scale factor between fits performed at twelve additional snapshots,
equally spaced in $z$, to $z = 0$. In all fits, we bound the total mass in the potential to be less than
$0.1$~dex greater than the stellar mass of the disk in the baryonic simulation. This method
yields an excellent match between the stellar mass in the baryonic simulations and the mass in
the embedded disks within fixed physical radii.\footnote{At the thirteen steps that anchor the
time evolution, the disk potential matches the stellar mass within 20~kpc to within $10\%$ in
\mf\ and within $20\%$ in \mi\ at worst; the average agreement over those thirteen steps is
$\sim7\%$ across both simulations.}

Table~\ref{tab:diskprops} lists the physical parameters of the analytic disk potential, and
equivalent quantities for the galaxies that form in the FIRE simulations.  The first two
columns quote the mass, quantified as $\Md$ for the disk potential (the total mass when integrated
to infinity) and by $M(<r_{90})/0.9$ for the FIRE simulations, where $r_{90}$ is the 3D radius
that contains 90\% of the mass.  The last two columns give the radial extent of the disk,
quantified here by the 2D radius $R_{90}$, where $M(R<R_{90}) = 0.9 \times M(z<z_{90})$ and
$z_{90}$ similarly contains 90\% of the total stellar mass of the galaxy.

While we carefully design the analytic potential to match the parameters of the stellar disk in
the baryonic simulations, we note several limitations. For simplicity, we fix the orientation
of the disk to be along the $x-y$ axis plane (rather than allowing it to rotate through
arbitrary angles).  Additionally, in the baryonic simulations, \mi\ and \mf\ do not form permanent
well-ordered disks until $z \approx 0.5$ and $z \approx 0.6$, respectively, so our assumed potential
overestimates the thinness of the disk at early times.  However, as we show below, the `thinness' of
the disk is not important for capturing its tidal effects on subhalos -- in fact, replacing the disk
with a \textit{spherical} \citet{Hernquist1990} profile of the same mass produces nearly the same
effect (see \S\ref{ssec:params}). Furthermore, we add this disk potential to the DMO simulation without
adjusting the mass of dark matter particles, thus slightly increasing the overall mass within the host
halo.  However, this error is likely small to the overall system because the disks comprise only
$\sim6\%$ of the total mass within $\rvir$ and $10\%$ within 80 kpc.  

We also fit our fiducial disks to only the stellar mass from the baryonic simulation; we 
do not try to fit the gaseous component.  This is a reasonable approximation at $z = 0$, where 
the gas fractions in the disks of our simulated host galaxies are $\approx 15\%$, comparable to 
the MW, M31, and similar galaxies \citep{Yin2009,Catinella2010,Catinella2012,Catinella2013}.  At 
higher redshifts, however, when the disks were more gas-rich, our method underestimates the total 
baryonic mass within the disk.  At these redshifts, stellar feedback drives significant gas flows 
in and out of galaxy on short timescales ($\lesssim 100$ Myr), compromising the accuracy of any 
simple, constant analytic description of the resultant potential. In principle, this lack of 
incorporating gas implies that our stellar-disk-only model represents a lower-limit to the level 
of substructure depletion from the central galaxy, though, as we show in \S\ref{ssec:params}
by increasing the mass of the embedded disk to include the gas in the galaxy, the purely gravitational 
contribution from gas is small.

However, as we demonstrate below, even this simple model is remarkably successful at reproducing the
statistical properties of surviving subhalos as compared with the fully baryonic simulations. Thus,
while there is room for further progress, our method represents a \textit{substantial} improvement
over DMO simulations at essentially the same CPU cost.

\begin{table}
\centering
\begin{tabular}{rlcc|cc}
    & redshift   &   $\Md$    &   $M_\mathrm{gal}$    &   $R_\mathrm{90,\,disk}$     &      $R_\mathrm{90,\,gal}$ \\
    &            &  $10^{10}M_\odot$    & $10^{10}M_\odot$                   &   $\kpc$    &   $\kpc$ \\
\hline\hline
\multirow{5}{*}{\mi} &
    $z = 3$ & 0.09 & 0.12 & 15.75 & 14.86 \\
    & $z = 2$ & 0.32 & 0.29 & 15.84 & 12.57 \\
    & $z = 1$ & 1.59 & 1.44 & 7.94 & 10.76 \\
    & $z = 0.5$ & 4.28 & 3.91 & 4.22 & 5.93 \\
    & $z = 0$ & 8.56 & 7.70 & 6.43 & 9.49 \\
\hline
\multirow{5}{*}{\mf} &
    $z = 3$ & 0.15 & 0.16 & 9.33 & 7.70 \\
    & $z = 2$ & 0.71 & 0.73 & 6.84 & 7.43 \\
    & $z = 1$ & 2.40 & 2.18 & 4.11 & 7.05 \\
    & $z = 0.5$ & 4.86 & 4.41 & 4.31 & 9.01 \\
    & $z = 0$ & 10.41 & 9.39 & 7.22 & 10.48 \\
\end{tabular}
\caption{
    Properties of the analytic `embedded disk' of the MW-mass central galaxies at several
    redshifts, along with the properties from the FIRE baryonic simulations that we model
    the disks on.  The first two columns list the total mass in the embedded disk, $\Md$,
    and the equivalent stellar mass in the FIRE galaxy; the following two columns list the radial
    extent of the disk (see text for details).  Our analytic disks are typically slightly
    more compact, particularly at late times, both because the Miyamoto-Nagai potential is
    an imperfect fit to the galaxies that form in the baryonic simulations, and because we
    fit the density profile, rather than the mass or the potential. As we show in \S\ref{ssec:params},
    the amount of substructure depletion most strongly depends on the disk mass; doubling
    $\Rd$ only weakly affects the surviving subhalo population.
}
\label{tab:diskprops}
\end{table}

\begin{figure*}
    \centering
    \includegraphics[width=\columnwidth]{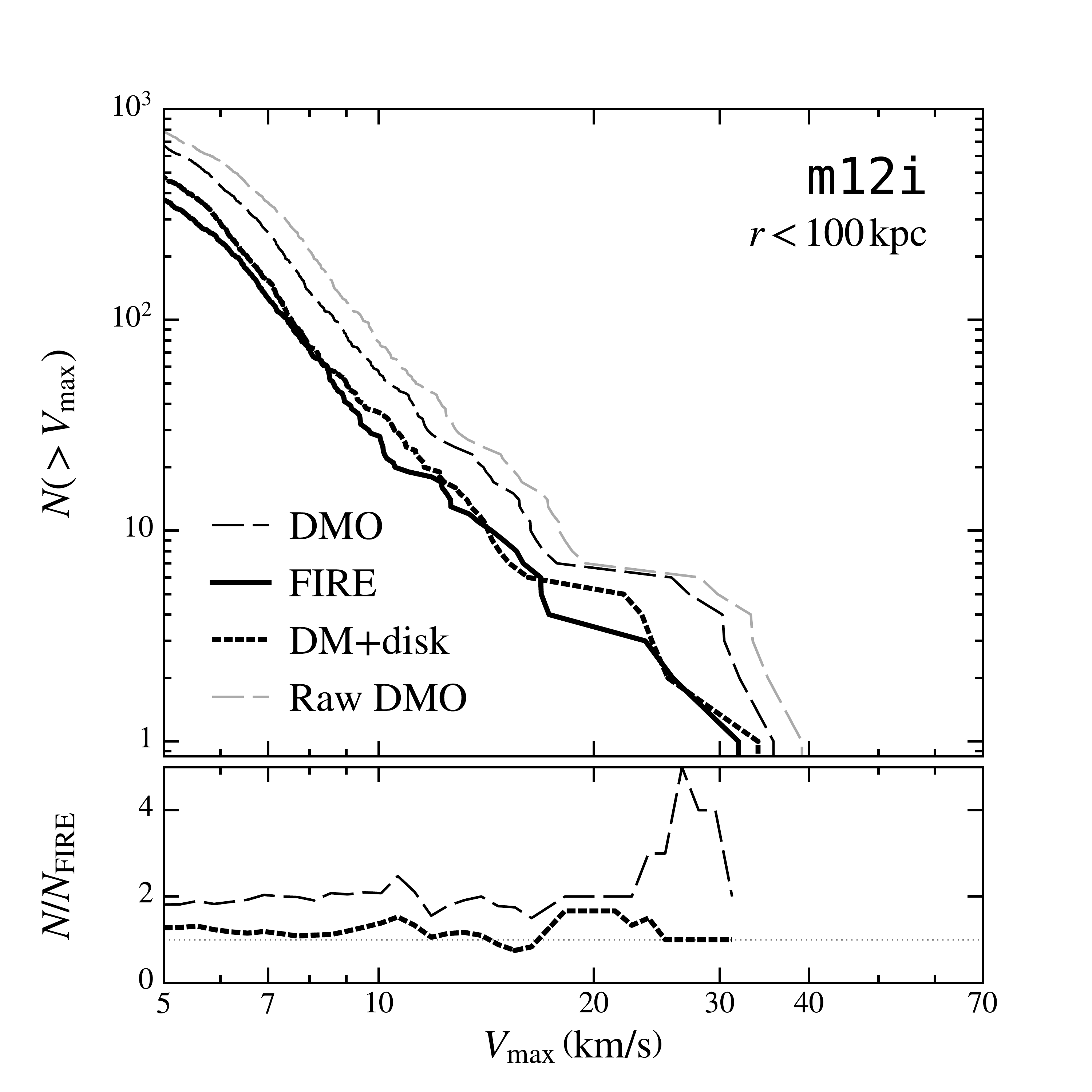}
    \includegraphics[width=\columnwidth]{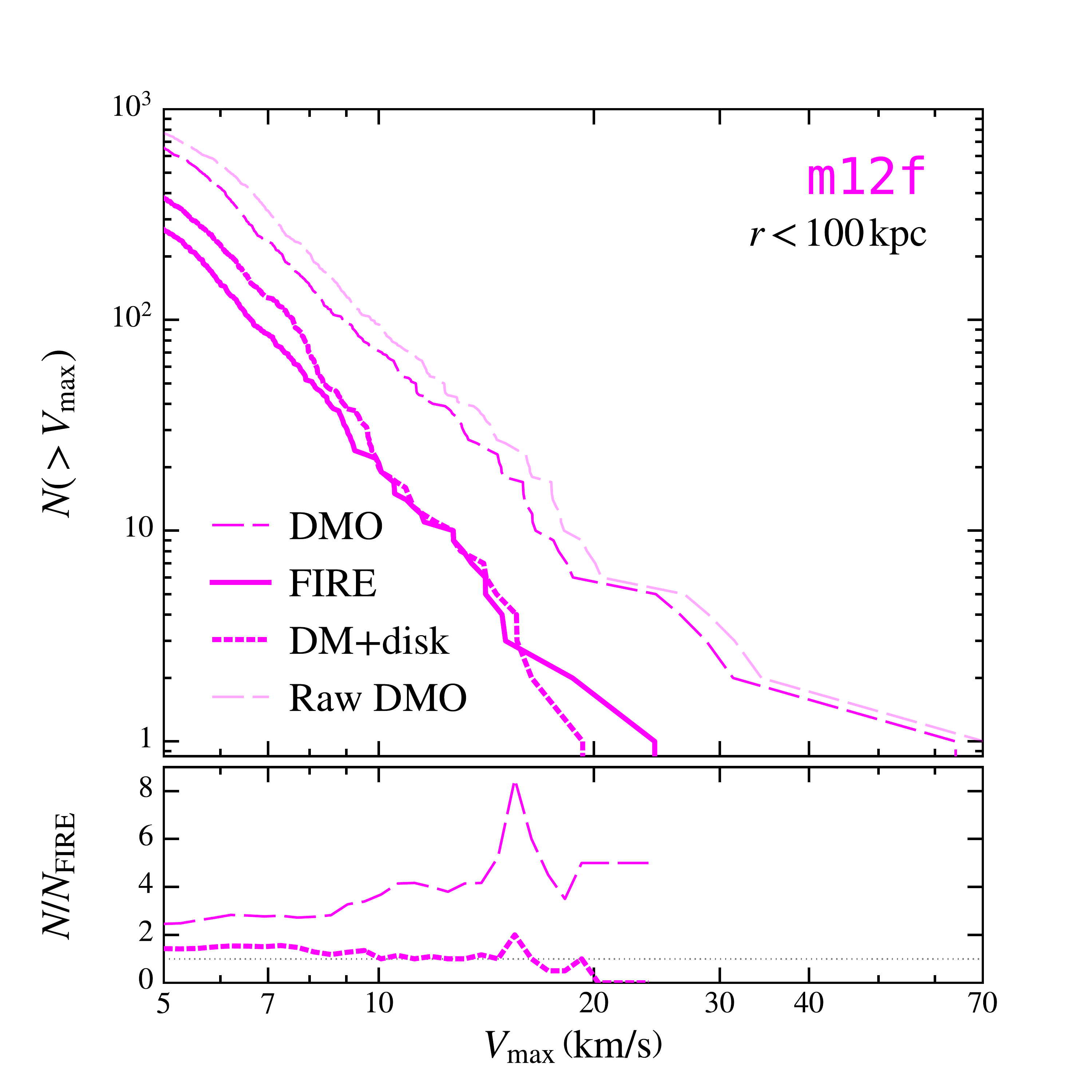} \\
    \includegraphics[width=\columnwidth]{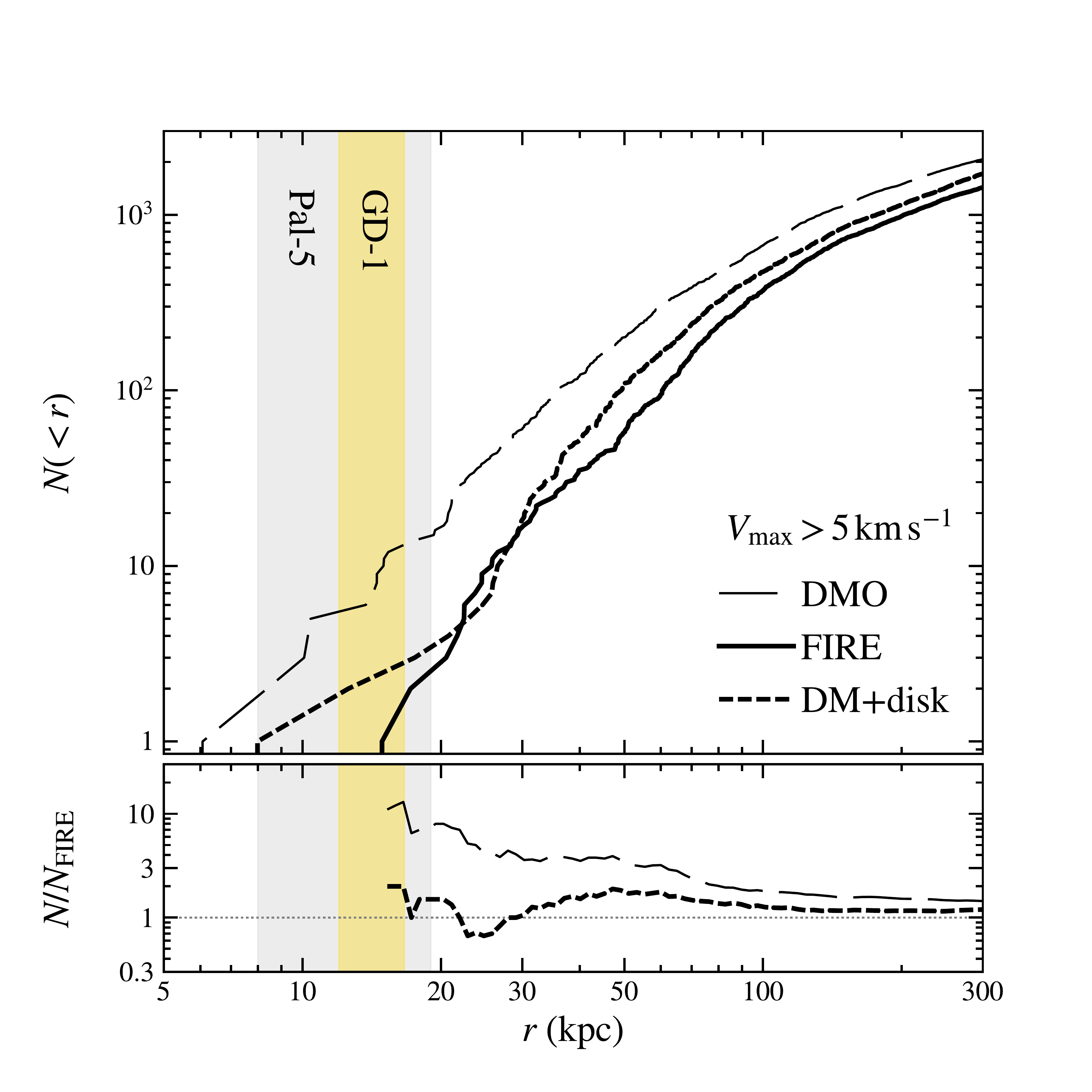}
    \includegraphics[width=\columnwidth]{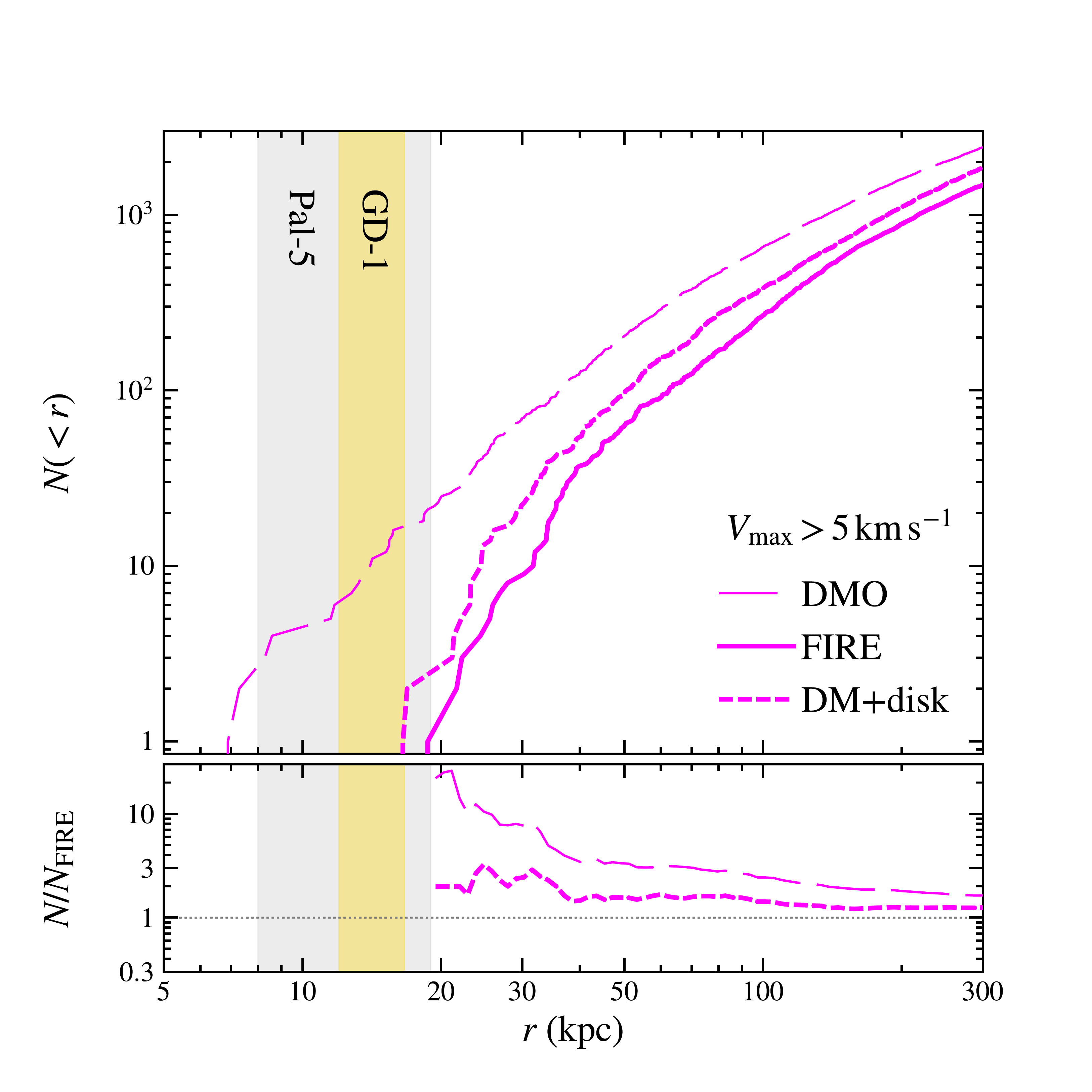}
\caption{
\textit{Top:} Cumulative counts of subhalos above a given maximum circular velocity, $\vmax$,
    within 100~kpc of the two hosts, \mi\ (\textit{left}) and \mf\ (\textit{right}) -- Appendix
    A presents the counts within 50~kpc and 300~kpc.  For reference, the upper-most, light colored
    dashed lines (labeled ``Raw DMO'') indicate the results of the DMO simulations \emph{without}
    applying the correction for the baryon fraction (that is, without multiplying by $\sqrt{1-f_\mathrm{b}}$).
    Henceforth we apply this correction for all comparisons.  Lower panels plot the ratio between the
    cumulative counts of subhalos in the DMO or embedded disk runs to the FIRE baryonic simulations.
    For both systems, the DMO simulation overpredicts the number of subhalos as compared with the
    baryonic simulation by at least $2 \times$ at all $\vmax$:  the average ratios plotted in the lower
    panels are $2.2$ and $3.9$ in \mi\ and \mf, respectively. Adding \emph{only} the galactic disk
    potential brings the substructure counts to within $\sim20\%$ agreement at all $\vmax$
    (average ratios of $1.2$ and $1.06$).
    \textit{Bottom:} Cumulative counts of subhalos within a given radius.  We include subhalos down
    to $\vmax = 5\,\kms$ (bound mass $M \simeq 5 \times 10^6 \Msun$), which are well-resolved.
    While the total excess of subhalos within $300\,\kpc \approx R_{\rm vir}$ is $\approx 50\%$ in the
    DMO simulations, this excess rises to $\approx 3 \times$ within 50 kpc.  Moreover, the disk
    \textit{completely} destroys all subhalos that are within $17$--$20~\kpc$ by $z = 0$, 
    where searches for dark substructure through stream heating are most sensitive: the light grey 
    and gold bands show the extent of the galactocentric orbits of Palomar-5 \citep{Carlberg2012} and 
    GD-1 \citep{Koposov2010}, respectively, the best-studied streams around the MW.  The embedded disk 
    simulations model this reduction/destruction to within a factor of 2 at all radii.  As we
    show in \S\ref{ssec:destroyed}, $\sim20$ subhalos with pericentric distances $<20~\kpc$ survive to 
    $z=0$.
}
\label{fig:vmaxdist}
\end{figure*}

\section{Results}
\label{sec:results}

Figure~\ref{fig:vizes} shows images of dark matter in our three types of simulations,
highlighting the qualitative effect of the central galaxy on the dark matter distribution
relative to the DMO simulation.  The top panels visualize, from left to right, the dark
matter density within a cube 500~kpc on a side in the DMO, FIRE baryonic, and embedded
disk simulations of \mi; the lower panels show a zoomed-in view of a cube 100~kpc across.
The log-scaled colormap changes from the upper to lower panels, but is identical across
the simulations. The dark matter near the central galaxies (lower panels) clearly responds
to the extra mass present in both the baryonic and the embedded disk simulations. However,
the dark matter takes on a slightly more disk-like shape around the analytic embedded potential;
we posit that the relatively spherical cusp around the central galaxy in the baryonic
simulation arises because of the continually evolving orientation of the central stellar
disk, along with the fluctuating gas distribution, which is more spherical in a time-average
sense because of time-dependent gas inflows and outflows.

Even more striking than the enhancement in the central dark matter density, however,
is the \textit{severe} reduction in the number of subhalos within the central
$r \lesssim 50~\kpc$. The central galaxy has destroyed an enormous fraction
of the satellites that the DMO simulation predicts in the central regions, where
observational probes are most sensitive.

\subsection{Dependence on mass and radius at $z = 0$}

Figure~\ref{fig:vmaxdist} quantifies the differences in the subhalo populations.
The top row shows the cumulative counts of subhalos within $100~\kpc$ of \mi\ (left)
and \mf\ (right) as a function of $\vmax$. The lower sub-panels show the ratio between
the cumulative counts in the collisionless simulations (with and without the embedded
potential) and the FIRE baryonic simulations. The DMO simulations, plotted in thin dashed
lines, overpredict the subhalo $\vmax$ function relative to the baryonic simulations,
plotted as solid lines, at all $\vmax$.  The light dashed lines, which plot counts in the
DMO simulations without correcting for the difference in particle masses between DMO and
baryonic simulations, show that DMO predictions improve relative to the FIRE simulations
after correcting for this effect, but only slightly -- the disk reduces counts at fixed
$\vmax$ by much more, particularly in \mf.  In fact, the embedded disk simulations provide
a significantly better match to the FIRE baryonic simulations at all $\vmax$. As demonstrated
in the bottom panels,
the disk simulations agree with the baryonic simulations to within $\sim20\%$ at nearly all
$\vmax$.

\begin{figure*}
    \includegraphics[width=\columnwidth]{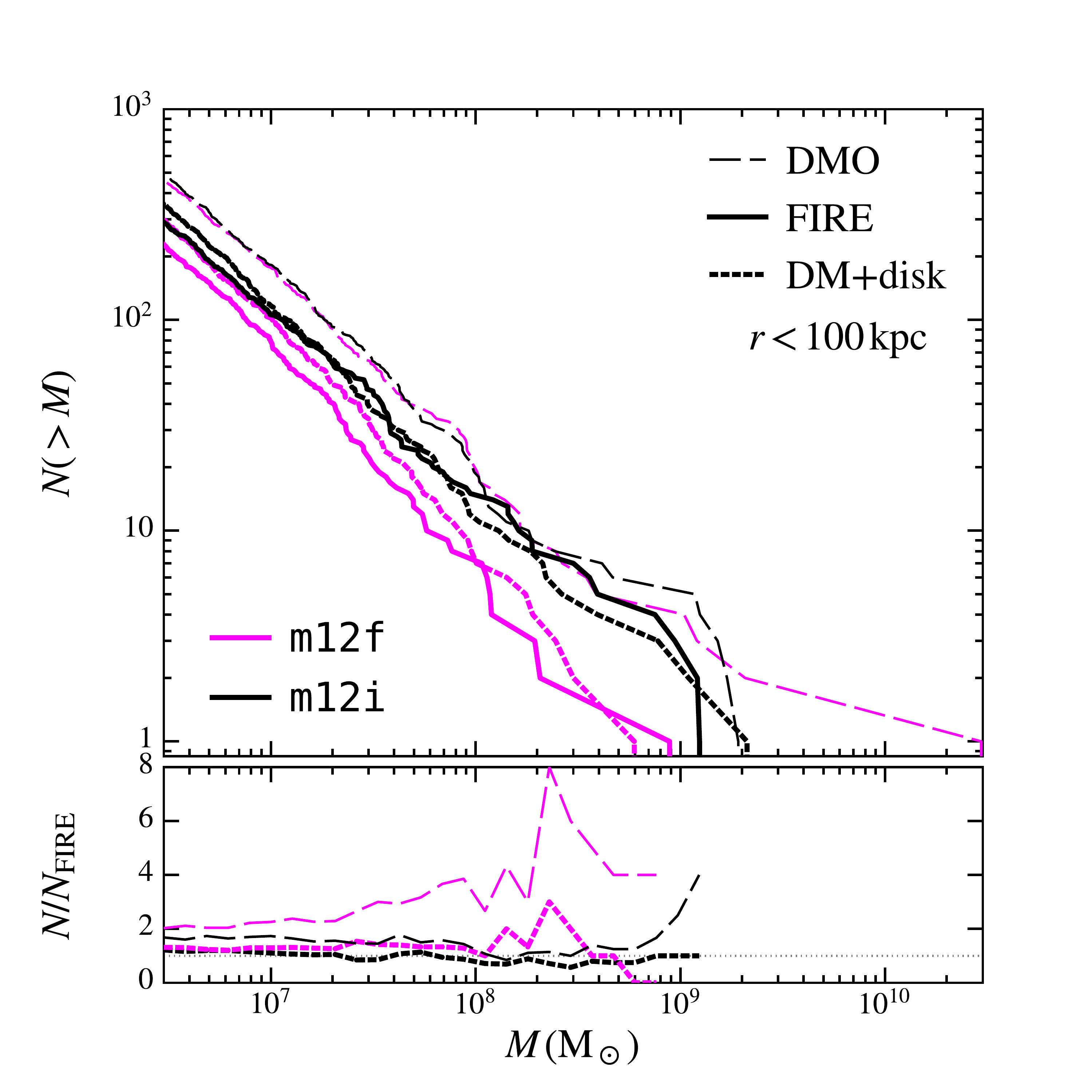}
    \includegraphics[width=\columnwidth]{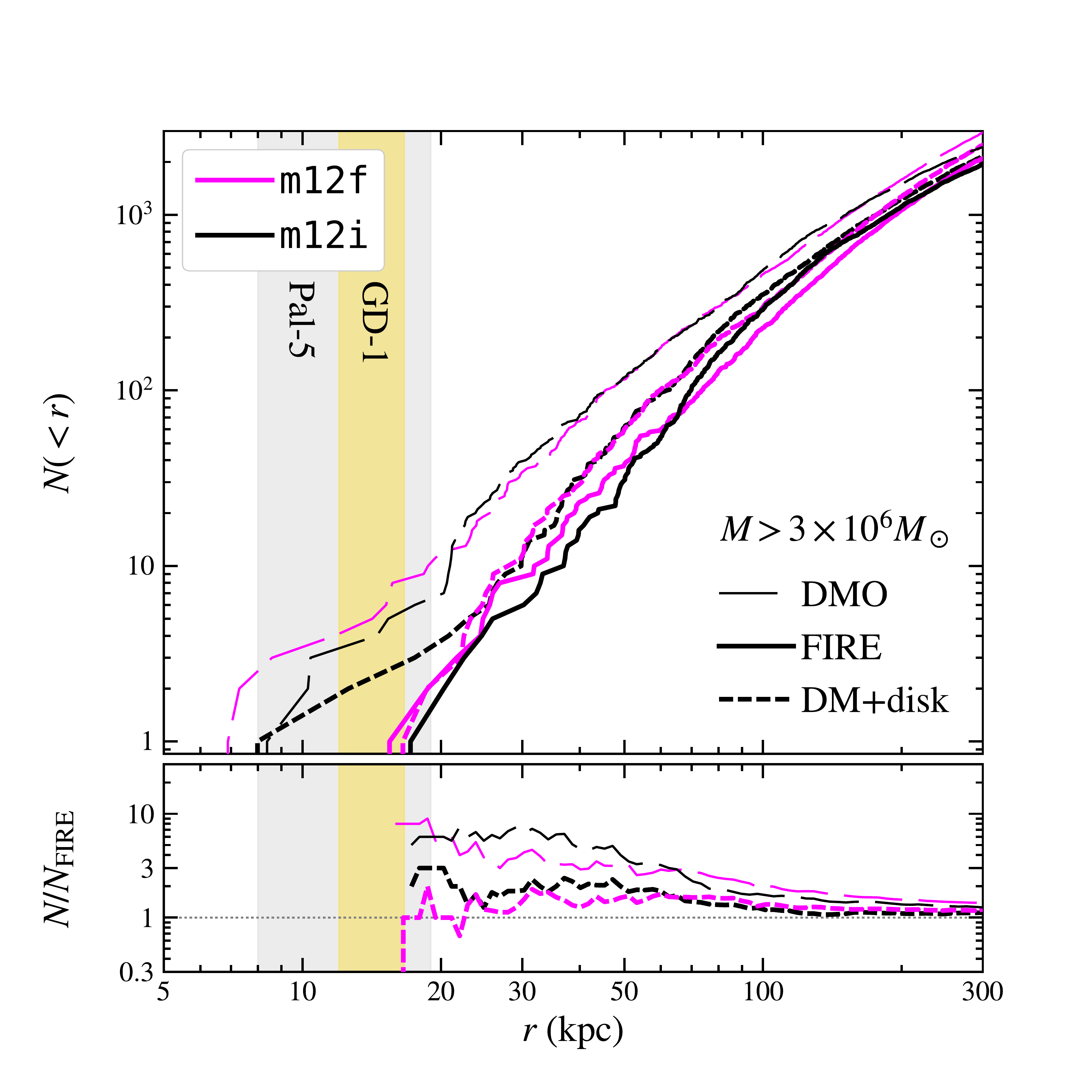}
\caption{
    Cumulative number of subhalos above a given bound dark matter mass, $M$,
    within $100~\kpc$ ({\em left}) and as a function of radius, $r$, for
    $M > 3 \times 10^{6} \Msun$ ({\em right}).  Lower panels plot the ratio
    between the cumulative counts of subhalos in the DMO or embedded disk runs
    relative to the FIRE baryonic simulations.  Selecting subhalos via $M$ instead
    of $\vmax$ (as in Figure~\ref{fig:vmaxdist}) leads to nearly identical results:
    DMO simulations overpredict the number of subhalos as compared with the baryonic
    simulations at all $M$ and $r$, while the embedded disk simulations show much better
    agreement.
}
\label{fig:massfuncs}
\end{figure*}

The bottom row in Figure~\ref{fig:vmaxdist} illustrates the radial
distribution of the subhalos included in the top row by showing the cumulative
number of subhalos as a function of 3D distance from the halo center, $r$. As
expected, the depletion is greatest at the center of the host: both the
\mi\ and \mf\ baryonic simulations have \emph{no} subhalos within $15~\kpc$ and
only $1$--$2$ within $20~\kpc$ at $z = 0$, while the DMO simulations predict
$\sim7$ subhalos within $15~\kpc$ and $\gtrsim 10$ within $20~\kpc$ (also see
Table~\ref{tab:sims}).  The embedded potential captures this effect well,
particularly around \mf: only two subhalos remain within $20~\kpc$. The match is
slightly worse around \mi, where three subhalos remain within $20~\kpc$ and two
within $15~\kpc$, but the embedded disk simulation still improves upon the DMO
simulation by a factor of $\sim3$.  Though we do not explicitly plot it here, we
also note that, even after accounting for the increased destruction via the disk,
the radial distributions of the surviving subhalos are more extended than predicted
from DMO simulations.

This dearth of substructure within $\sim20~\kpc$ has strong implications for attempts to discover
dark subhalos via stellar streams around the MW. Thus far, the best studied streams, Palomar 5 at
$19~\kpc$ \citep[indicated by the grey vertical band;][]{Carlberg2012} and GD-1 at $\sim15~\kpc$
\citep[plotted as the gold band;][]{Koposov2010}, are within this distance, suggesting that baryonic
simulations predict far fewer gaps in stellar streams than DMO simulations. The null result from
\citet{Ibata2016} therefore may be in line with predictions from \lcdm\ \textit{including baryonic
effects}.  However, those gaps could be created by subhalos at an earlier point in their orbits
than their positions at $z = 0$; we therefore examine the distribution of pericentric distances
$d_\mathrm{peri}$ in \S\ref{ssec:destroyed}.

While $\vmax$ can be measured robustly in simulations, the degree to which
a subhalo disrupts a stellar stream is more closely related to its bound mass
\citep[along with the interaction distance and velocity; e.g.][]{Yoon2011,Carlberg2013,Sanders2016,Sanderson2016}.
Therefore, the left panel of Figure~\ref{fig:massfuncs} plots the cumulative
subhalos count, again within $100~\kpc$, as a function of the bound mass assigned
to them by \texttt{AHF}, $M$. Because we run \texttt{AHF} only on dark matter particles,
$M$ therefore represents the bound dark matter mass. The results are similar to those using
$\vmax$ as in Figure~\ref{fig:vmaxdist}, though the discrepancies between the DMO and
baryonic simulations are less severe: the DMO simulations overpredict subhalo
counts above fixed $M$ by a factor of $\gtrsim 1.7$.

The right panel of Figure~\ref{fig:massfuncs} plots the cumulative
radial distribution of subhalos with $M > 3 \times 10^6 \Msun$, which
are reliably resolved in our simulations with 85 particles (Appendix~B presents
an explicit resolution test). The distributions are again similar to those in the
lower plots in Figure~\ref{fig:vmaxdist}: simulations with a central galaxy predict
only $1$--$3$ subhalos in the central $20~\kpc$, while the DMO simulations overpredict
that count by a factor of $> 5$. Therefore, the lack of substructure near the galaxy is
independent of whether subhalos are selected by $\vmax$ or $M$.

Because correcting for the reduction in the particle mass from the lack of baryons
is more straightforward for $M$ than for $\vmax$, which depends on the mass
\emph{profile} of the subhalo, we select subhalos based on $M$ for the remainder of
the paper, using the resolution cut of $M > 3 \times 10^6 \Msun$. 

In general, the differences between the DMO and FIRE baryonic simulations are
the largest among the most massive subhalos ($\vmax \gtrsim 25~\kms$),
which form a non-negligible amount of stars and are therefore influenced
by both internal and external baryonic effects. These are the subhalos
that are important for the too-big-to-fail problem, and the relative lack of
these subhalos in the baryonic simulations is key for solving the
too-big-to-fail problem in the satellite populations of those hosts. Surprisingly,
even without including the internal changes driven by feedback and bursty outflows,
the embedded disk simulations capture these quantitative trends remarkably well: the
accuracy of the embedded disk simulations is largely independent of subhalo mass at
$r \lesssim 100~\kpc$. This agreement suggests that subhalo survival near the center
of the host halo is dominated by gravitational interactions, such that these subhalos
are likely to suffer the same fate of disruption independent of their internal structure
(on $\sim500~\pc$ scales). However, when including all subhalos out to $300~\kpc$ (see
Appendix~A), the disk simulations do overpredict the number of subhalos with
$\vmax \gtrsim 25~\kms$ by a factor of $3 - 5$ relative to the baryonic simulations,
indicating that internal baryonic processes remain important in regulating the
population of subhalos at larger distances.  These results generally agree with those of
\citet{BrooksZolotov2012}, who found that tidal effects dominate for subhalos that
pass near the galactic disk.

Therefore, the tidal field from the central galaxy is partially responsible
for eliminating some of the subhalos that a too-big-to-fail analyses would
identify as problematic `massive failures' in a DMO simulation. As shown in
\citet{Wetzel2016}, \mi\ does not suffer from a too-big-to-fail problem
when simulated with FIRE baryonic physics, indicating that a combination of
internal feedback and tidal interactions resolve TBTF around that galaxy.
The embedded disk potential alone reduces the number of subhalos with
$\vmax > 25~\kms$ within $300~\kpc$ from $12$ to $7$, suggesting that the central
galaxy is responsible for roughly half of the necessary changes \emph{for this system},
with internal feedback driving lower inner densities in the surviving subhalos and
accounting for the remaining discrepancy. We draw similar conclusions from \mf, which
hosts $9$ subhalos with $\vmax > 25~\kms$ in the DMO run, but only $5$ when simulated
with an embedded disk, and identically zero when simulated with full physics. Of course,
the exact relative contribution from internal feedback versus external tidal forces depends
strongly on the individual histories of those large subhalos, their spatial distribution
within the host and, as we will show below, their orbital characteristics.

\subsection{Destruction versus mass stripping?}

\begin{figure}
\centering
    \includegraphics[width=\columnwidth]{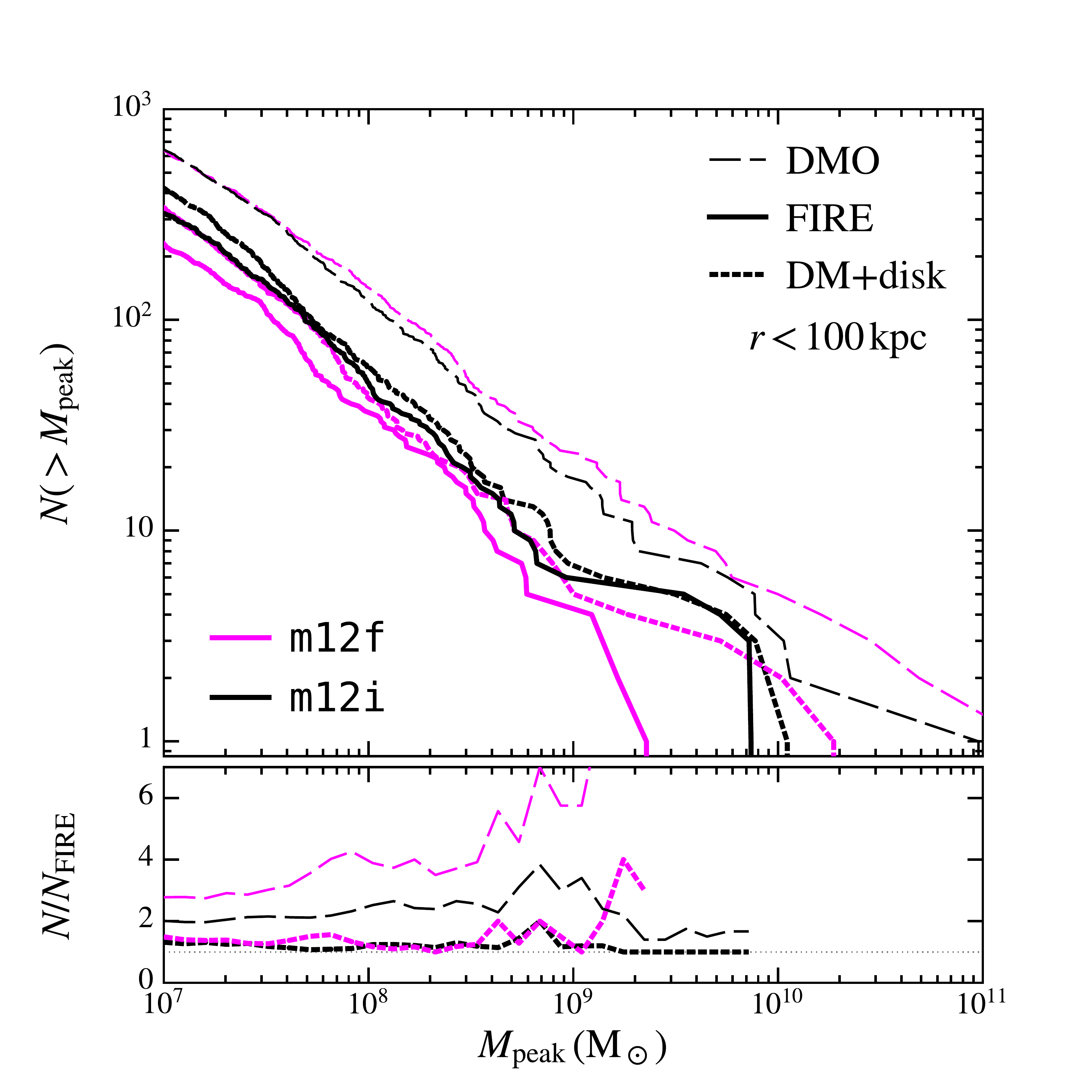}
\caption{
    Cumulative number of subhalos within $< 100\,$kpc at at $z = 0$, as in
    Figure~\ref{fig:massfuncs}, but as a function of $\mpeak$, the largest
    virial mass ever attained by each subhalo (before any stripping).
    Though the agreement between the embedded disk and FIRE baryonic
    simulations are marginally worse than in Figures~\ref{fig:vmaxdist} or
    \ref{fig:massfuncs}, which measured \textit{instantaneous} $\vmax$ or bound
    mass $M$, the former still provide a better match to the FIRE simulations
    than the DMO predictions.
    Because $\mpeak$ is sensitive to destruction but is insensitive to (partial)
    mass stripping, this agreement indicates that, relative to the DMO simulations,
    the galactic disk primarily destroys subhalos and removes them from the population
    -- if subhalos were stripped after infall, but survived until $z=0$, then counts
    in the simulations with a central galaxy would agree nearly perfectly with those
    in the DMO simulations.
}
\label{fig:peakcounts}
\end{figure}

In general, the tidal force from the central disk potential can either strip off a portion
of the outer mass of a subhalo, shifting it to lower $M$ and $\vmax$, or it can completely
destroy the subhalo, either through tidal shocking \citep{Gnedin1999} or repeated stripping
events, removing it from the population entirely. In order to distinguish between subhalo mass loss
from total destruction, we measure subhalos using $\mpeak$, defined as the largest virial
mass that the main branch of each subhalo's merger
tree reached. Thus, this quantity is preserved for a subhalo after infall and is not susceptible
to mass loss, provided the subhalo survives to $z = 0$.

Figure~\ref{fig:peakcounts} shows the cumulative counts of resolved subhalos within $100~\kpc$
as a function of $\mpeak$.
If the central galaxy typically strips subhalos, but does not entirely destroy them, then the
$\mpeak$ functions from the FIRE and disk simulations will agree with those of the DMO runs.
Instead, at fixed $\mpeak$, the number of subhalos is at least $2 - 3 \times$ larger in the
DMO simulation, meaning that the central galaxy has destroyed at least $50 - 70\%$ of
the substructure that currently resides within $100~\kpc$ of the halo center. As before, the
embedded disk captures all but $\sim10$--$20\%$ of this destruction, independent of
$\mpeak$ except for the few largest subhalos in \mf.

\subsection{Which subhalos are destroyed?}
\label{ssec:destroyed}

\begin{figure*}
\centering
    \includegraphics[width=\columnwidth]{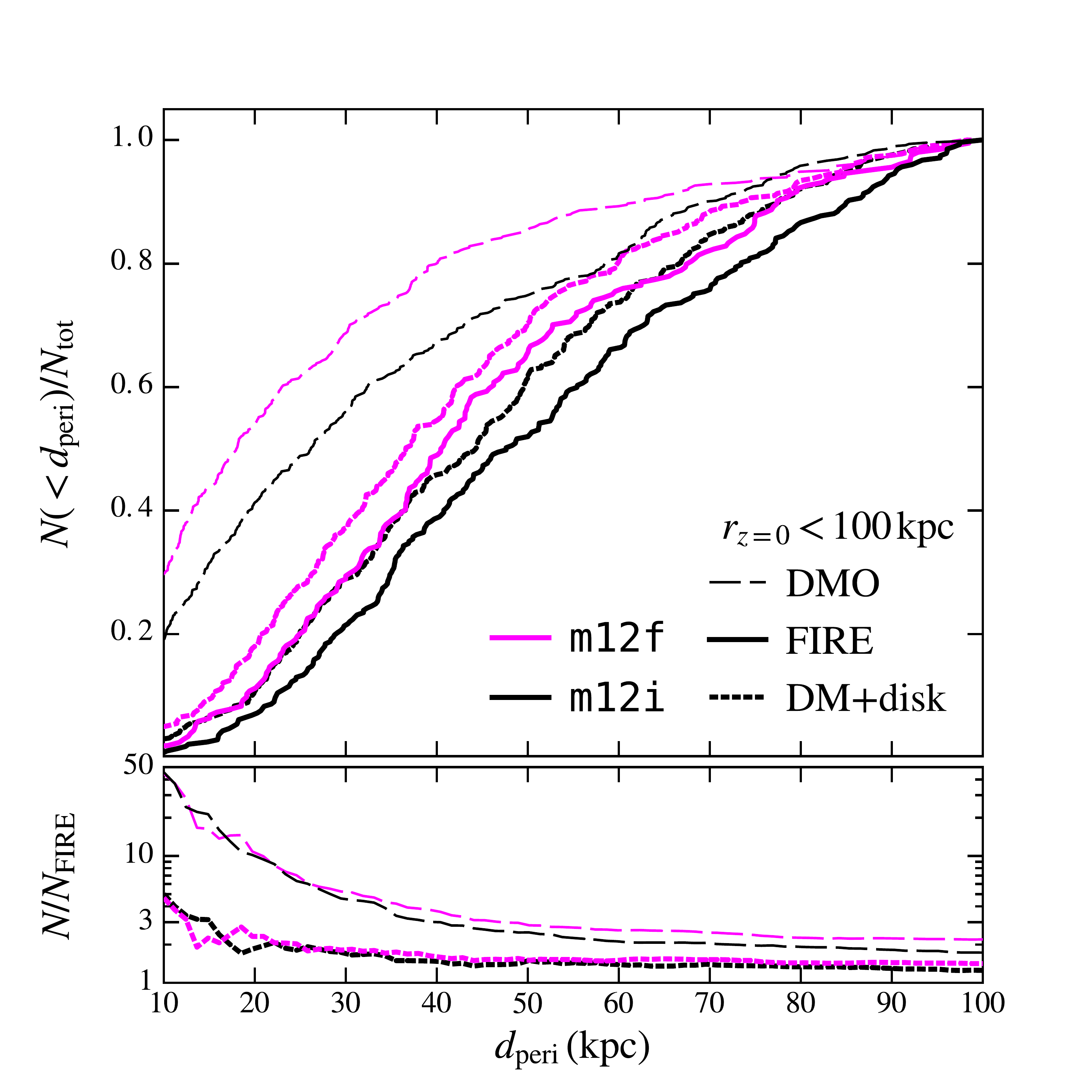}
    \includegraphics[width=\columnwidth]{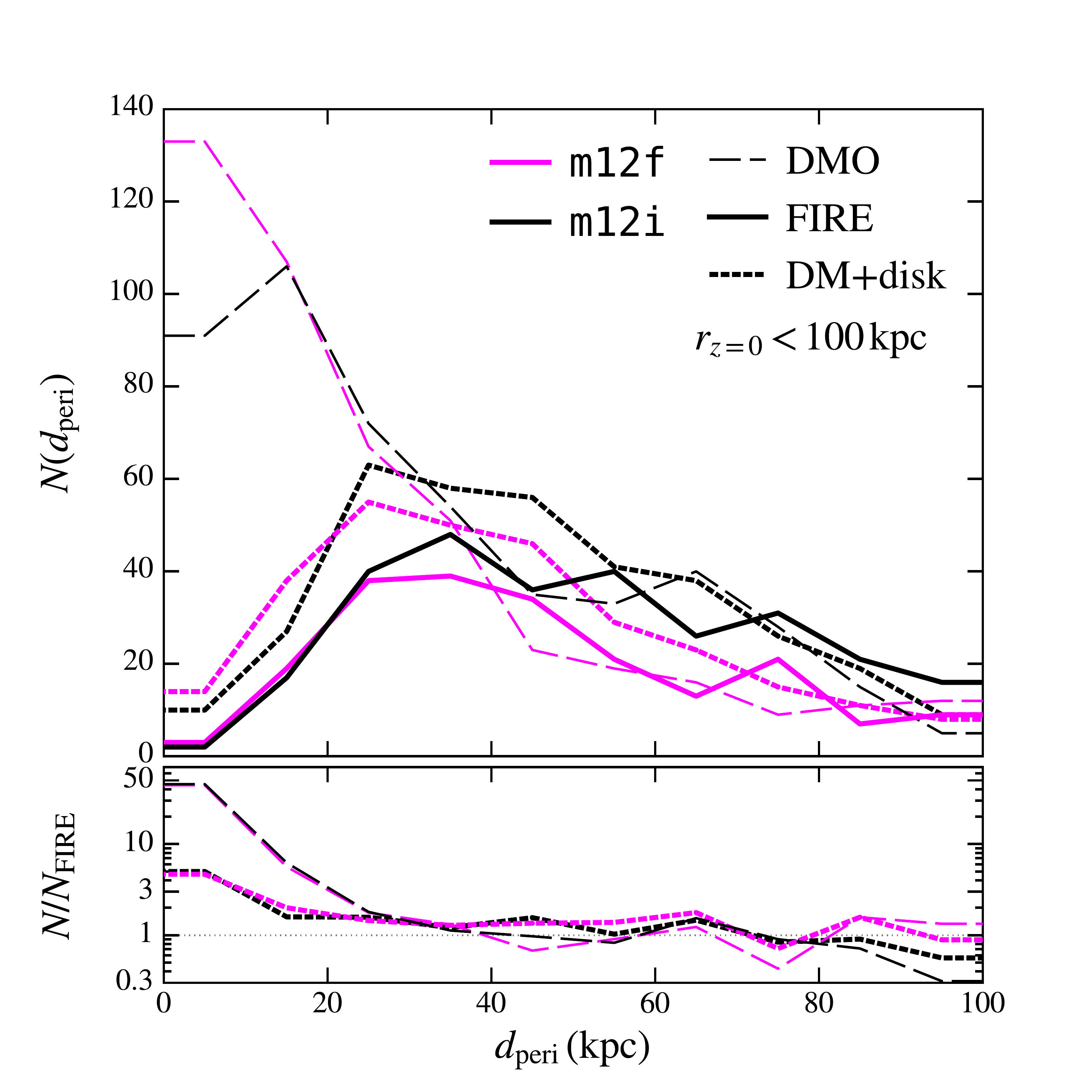}
\caption{
    Cumulative (\textit{left}) and differential (\textit{right}) counts of subhalos
    within $100~\kpc$ at $z = 0$ as a function of their distance of closest approach
    to halo center, $\dperi$.  The distributions in the left upper panel are normalized
    to the total number with $d_\mathrm{peri} < 100~\kpc$, while the left lower panel
    indicates fractional differences in the \emph{absolute} numbers; they are therefore
    not expected to equal 1 at $d_\mathrm{peri} = 100~\kpc$.  The distributions of
    pericentric distances in the DMO simulations far exceed those from the embedded disk
    or baryonic runs within $\dperi \lesssim 30~\kpc$, indicating that the central galaxy
    is responsible for subhalo destruction within that radius. DMO simulations overpredict
    the number of surviving subhalos that passed within $10~\kpc$ by a factor of $\sim50$,
    and within $20~\kpc$ by $\sim15 \times$.  The embedded disk simulations are accurate to
    within a factor of $\approx 2 - 4$. The excess at  $\dperi \lesssim 20~\kpc$ in the disk
    runs is at least partially caused by not including the gaseous contribution: doubling the
    mass of the disk (\S\ref{ssec:params}) eliminates nearly all of the excess around \mi. The
    median pericentric distances experienced by surviving subhalos in the simulations with a disk are
    $\sim2$ times larger than in the purely DMO simulations.  The stark differences between
    the DMO and embedded disk simulations in these counts, which are insensitive to subhalo mass,
    suggest that the primary physical effect of the disk potential, particularly for those that
    come within $\sim30~\kpc$, is to destroy subhalos and remove them from the population, rather
    than to partially strip their mass.
}
\label{fig:dperi}
\end{figure*}

Thus far, we have examined subhalos only as a function of their position at $z = 0$.
However, we also can ask: out to what orbital pericenter distance are subhalos
significantly affected by the presence of the central disk? The left panel of
Figure~\ref{fig:dperi} shows the cumulative distribution of the pericentric distance,
$\dperi$, that subhalos within $100~\kpc$ that survive to $z=0$ experienced, normalized
to the total number of subhalos within $100~\kpc$.  We define $\dperi$ as the smallest
physical distance reached between the main branch of the host and that of a given
subhalo.\footnote{The positions of the two halos are interpolated in scale factor with a
third-order spline to improve the time resolution, as in \citet{Fillingham2015}.}
The curves, which use the same colors and styles as previous Figures, are thus
required to equal $1$ at $100~\kpc$. The lower panel plots the ratio of the cumulative,
non-normalized distributions, again relative to the FIRE baryonic simulations. The right
panel of Figure~\ref{fig:dperi}, meanwhile, plots the (non-normalized) differential distribution
of these same subhalos.

Clearly, when a central galaxy is present, the orbit of a subhalo is important to
its survival, and the differences in the distributions are stark: the median pericentric
distances of the DMO samples are $20~\kpc$ smaller than when the disk is included.
Nearly every subhalo that reaches the central $10~\kpc$ is destroyed by the central galaxy:
only 2 (3) subhalos that have passed within $10~\kpc$ of the center of \mi\ (\mf)
survive to $z=0$ in the baryonic simulations, whereas $\sim100$ such objects exist in the DMO
runs. Similarly, $50\%$ of the surviving subhalos have pericentric distances
$\lesssim 20~\kpc$ in the DMO simulation of \mf; that fraction drops to $\sim5\%$
in both the baryonic and embedded disk simulations. The impact of the central galaxy
is much weaker for subhalos that have never passed within $\sim30~\kpc$.  This is
in qualitative agreement with previous simulations by \citet{DOnghia2010}, who reached
similar conclusions by inserting a disk potential into a halo extracted from a
cosmological simulation.

The embedded disk simulations match the baryonic simulations well: the median
pericentric distances are only $\sim5~\kpc$ smaller in the former, and the lower panels
demonstrate that the distributions match better than a factor of $2$ down to
$\dperi \approx 20~\kpc$, though the embedded disk does overestimate the number
of subhalos with $\dperi \sim15~\kpc$. The disk simulations remain an improvement
over purely DMO simulations, however: the latter over-predict the number of surviving
subhalos that have passed within $10~\kpc$ of a MW-like galaxy by a factor of $\sim50$.
The remaining excess in the disk simulations is at least partially caused by the reduced
central mass at early times due to our exclusion of gas from the disk model: doubling the
mass of the disk at all times (\S\ref{ssec:params}) completely eliminates this discrepancy
within $10~\kpc$, though such a change does overcorrect for the total mass in the FIRE
simulations after $z \sim1.5$.

The distributions plotted in Figure~\ref{fig:dperi} are largely independent of
bound mass $M$:  increasing the mass cut to only include subhalos with
$M > 3\times10^7\msun$ yields nearly identical trends.  Therefore, the sharp contrast
between the DMO and embedded disk simulations further suggests that the primary role
of the disk is to completely destroy subhalos and remove them from the population,
particularly for subhalos that come near to the galaxy.  In fact, the differences
in the count of resolved subhalos within $100~\kpc$ of both hosts can be completely
accounted for by the difference in the number of subhalos with $\dperi < 20~\kpc$.

The distributions in Figure~\ref{fig:dperi} also inform our understanding
of the differences between \mi\ and \mf. Specifically, the relative depletion in
\mf\ is larger than in \mi, likely because the pericenter distribution
in the former is skewed to smaller radii, making its subhalo population
more sensitive to the presence of a central galaxy.\footnote{While \mi\ and \mf\ have 
a similar total mass within a 2 Mpc sphere, \mf\ is within a more filamentary environment, 
which may explain why its satellites have more radial orbits.} This difference motivates the
need to run many such simulations to explore the full range of subhalo orbital
characteristics between hosts.

Figure~\ref{fig:vmaxdist} showed that \mi\ and \mf\ baryonic simulations have
\textit{no} resolved subhalos within $17 - 20 \,\kpc$ of the central galaxy at $z = 0$.
However, Figure~\ref{fig:dperi} shows that $\approx 20$ subhalos have orbited within
this distance at some point, they simply are not located within this region by $z = 0$.
Furthermore, Figure~\ref{fig:dperi} plots only pericentric distances of \emph{surviving}
subhalos, and those that have been destroyed also could have induced dynamical effects in
the inner halo, such as creating holes in stellar streams, prior to their destruction.
That said, our results show that the total timescales over which such subhalos could act
is significantly shorter. We leave the broader question of how streams form and evolve in fully
cosmological simulations that include a central galaxy for later work.

\begin{figure*}
\centering
    \includegraphics[width=\columnwidth]{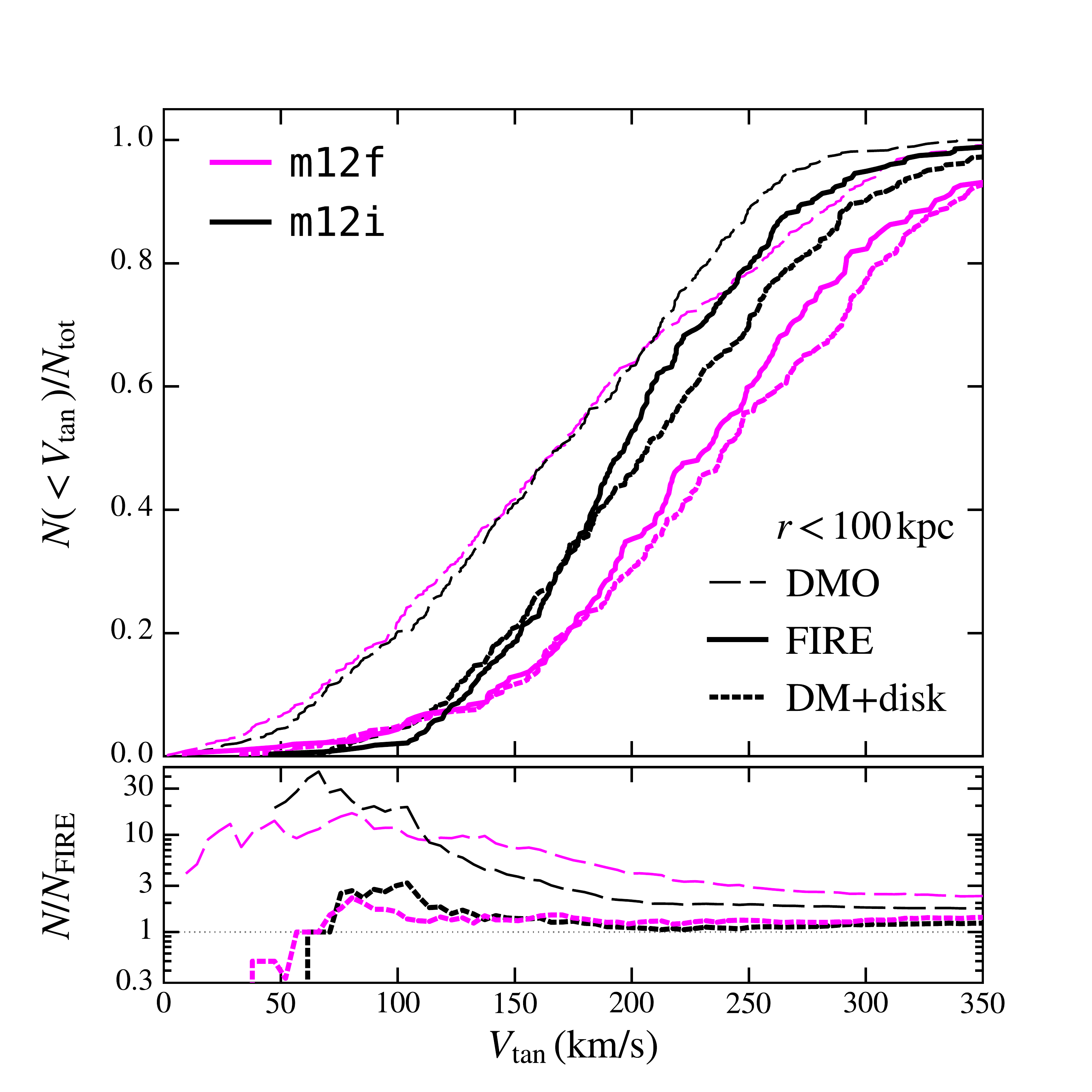}
    \includegraphics[width=\columnwidth]{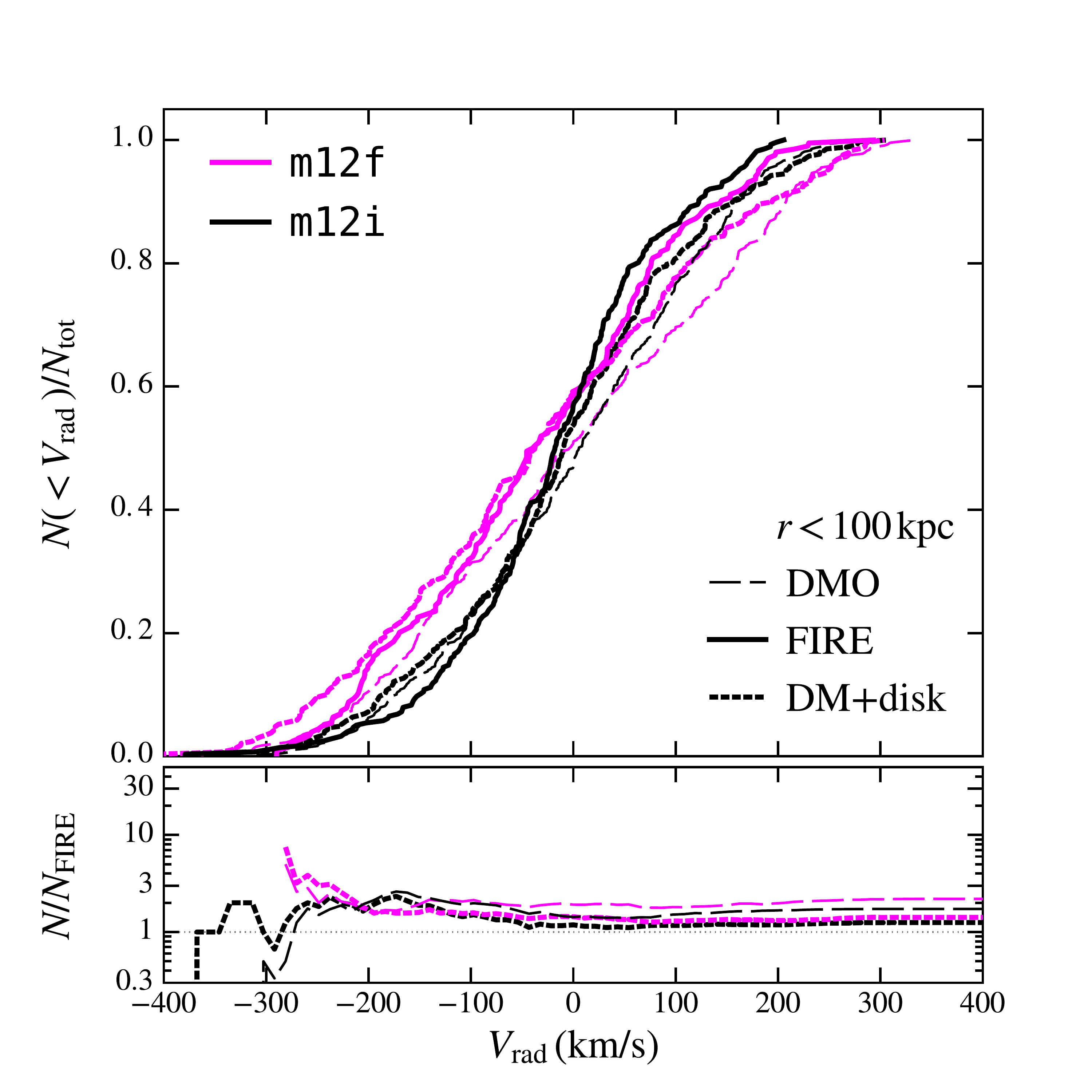}
\caption{
    Cumulative distribution of tangential (\emph{left}) and radial (\emph{right})
    velocities of resolved subhalos within $100~\kpc$ of the center of \mi\ 
    (black lines) and \mf\ (magenta lines) at $z = 0$, normalized to the total 
    number within $100~\kpc$. Consistent with the baryonic simulations, the embedded 
    disk preferentially destroys subhalos that are on radial orbits (low $\vtan$) that 
    pass close to the disk. Similarly, subhalos with high radial velocities 
    ($\vrad \gtrsim 75~\kms$) are slightly suppressed in the baryonic and embedded disk 
    simulations, but the difference is far less pronounced than in $\vtan$. As in 
    Figure~\ref{fig:dperi}, the lower panels shows the fractional difference in the absolute 
    (non-normalized) cumulative distribution.
}
\label{fig:velocities}
\end{figure*}

Figure~\ref{fig:dperi} clearly shows that $\gtrsim 90\%$ of subhalos that pass within
$10$--$20~\kpc$ of the central galaxy are destroyed. This dependence on pericenter implies
that the velocity distribution of \textit{surviving} subhalos should be significantly biased
relative to DMO simulations, such that systems on radial, plunging orbits with low specific
angular momentum should be preferentially destroyed.

The left panel of Figure~\ref{fig:velocities} shows the cumulative distribution of 
the tangential velocities, $\vtan$, of subhalos within $100~\kpc$, normalized to the 
total number within $100~\kpc$, demonstrating the strength of this effect. While subhalos 
in the DMO simulation have a mean $\vtan$ of $\sim150~\kms$, the presence of a central galaxy
increases it to $\sim~200-250~\kms$. The lower panel, which again plots the ratio in the
absolute distributions, demonstrates that the DMO simulations over predict the number of
subhalos within $100~\kpc$ with $\vtan < 100~\kms$ by a factor of ten. The embedded disk
simulations, however, agree to within a factor of $2$--$3$ at all $\vtan$.  While the
suppression of subhalos with low $\vtan$ in runs with a central galaxy is primarily
caused by the destruction of those systems, the deeper potential well in the central regions
also leads to an increase in the velocities of surviving systems near the galaxy.

We similarly examine the distributions of subhalo radial velocities in the right panel
of Figure~\ref{fig:velocities}. However, in contrast to the tangential distributions, we 
find that they agree across all simulations to within $\sim50\%$ at most $\vrad$. We do 
find that the relative fraction of subhalos with large, positive radial velocities 
($\vrad \gtrsim 75~\kms$) is slightly suppressed in the simulations with a central galaxy 
relative to the DMO simulations, as expected if subhalos that have passed close to the halo 
center and would be moving away from the host are preferentially destroyed.  However, overall 
these results indicate that \textit{a subhalo's likelihood of being destroyed by the central 
galaxy does not depend significantly on its radial velocity}. As a result, the primary effect 
of the central galaxy is on the resultant distribution of tangential, and not radial, velocities 
of the surviving subhalo population. There is a slight trend wherein the pure DMO 
simulations underpredict the fraction of high $\left|\vrad\right|$ subhalos (i.e., overpredict 
the fraction of systems with $\vrad\sim0~\kms$), consistent with an overall increase in 
subhalo velocities in the presence of a galaxy due to the added mass at the center of the host.


This tangential velocity bias may have direct consequences on the expected
perturbations induced in the cold stellar streams that are suggested as best able
to constrain dark subhalo counts in the MW, which stretch along the plane of the
sky \citep[e.g.][]{Newberg2016}. Specifically, if both the observed streams and
the subhalo population are tangentially biased, one would expect more `glancing'
blows, wherein subhalos are either co-rotating or counter-rotating
relative to the orbit of the stream, and fewer perpendicular interactions,
where the subhalo punches directly through a cold stream, relative to what is
expected for an isotropic population of subhalos \citep[the standard assumption in most
work on stream-subhalo interactions, e.g.][]{Yoon2011,Carlberg2012a,Erkal2016}. 
Counter-rotating interactions will have a small impact on streams due to their 
relatively short interaction times, but a larger fraction of the encounters may by 
co-rotating than otherwise expected. These occur over long enough timescales that the 
impulse approximation typically used to estimate changes in the stellar velocities 
\citep[e.g.][]{Sanders2016} may break down. We defer a full analysis of the degree 
to which this effect is observable to future work.

Similarly, the preferential destruction of subhalos on highly radial orbits has
important implications for the structure of the stellar halo, particularly close
to the MW. Specifically, the majority of the destroyed satellites that make up the
stellar halo \citep{Bullock2001,Bullock2005,Bell2008} were likely on plunging, highly radial orbits,
such that their resultant stellar streams are likely to be radially extended and reach
to greater radii. Therefore, streams from destroyed satellites may prove useful for
detecting clear, coherent gaps from the surviving dark subhalos, which are largely on
tangential orbits that are perpendicular to the expected orientations of the streams from
destroyed satellites.

\subsection{Varying the disk parameters}
\label{ssec:params}

\begin{figure*}
\centering
    \includegraphics[width=\columnwidth]{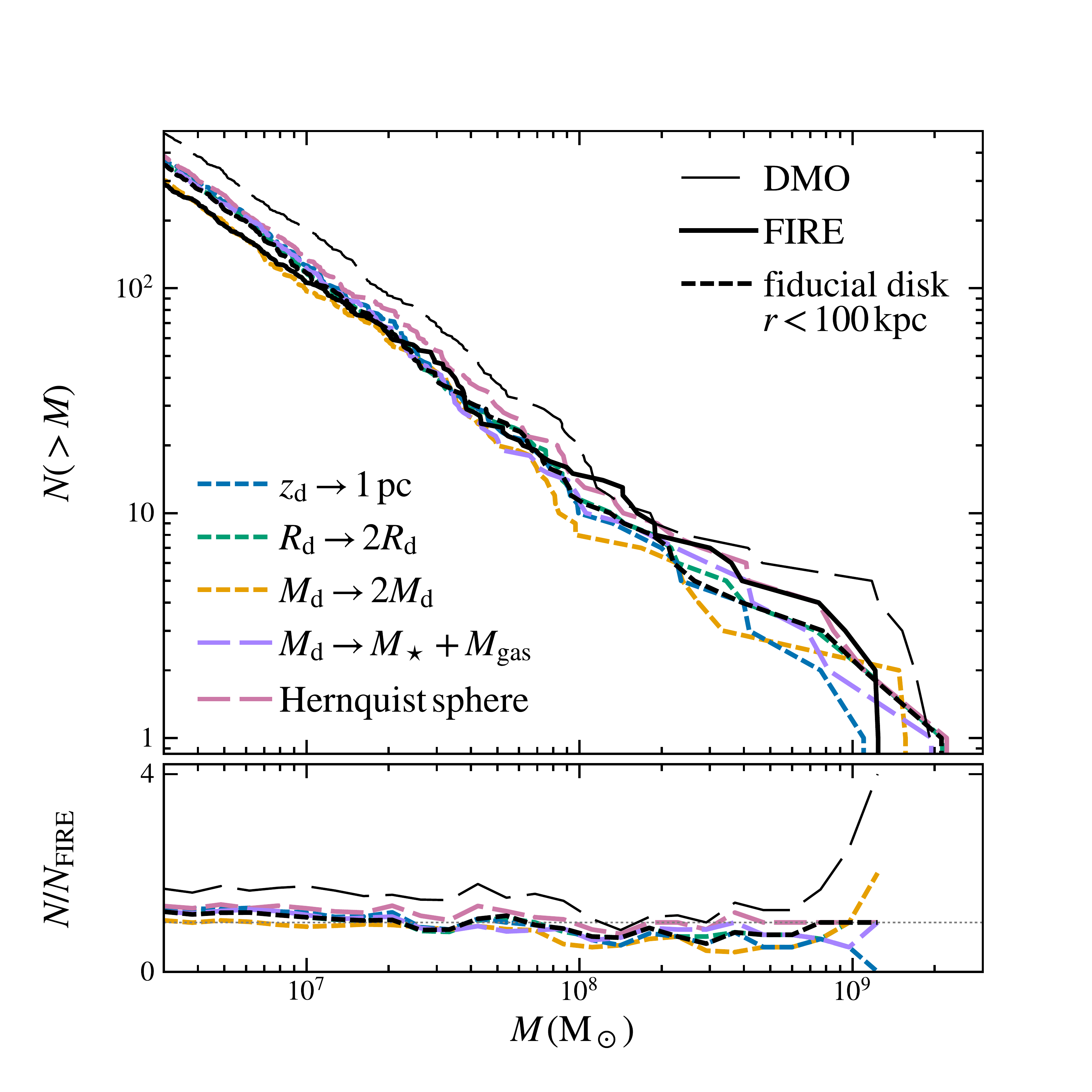}
    \includegraphics[width=\columnwidth]{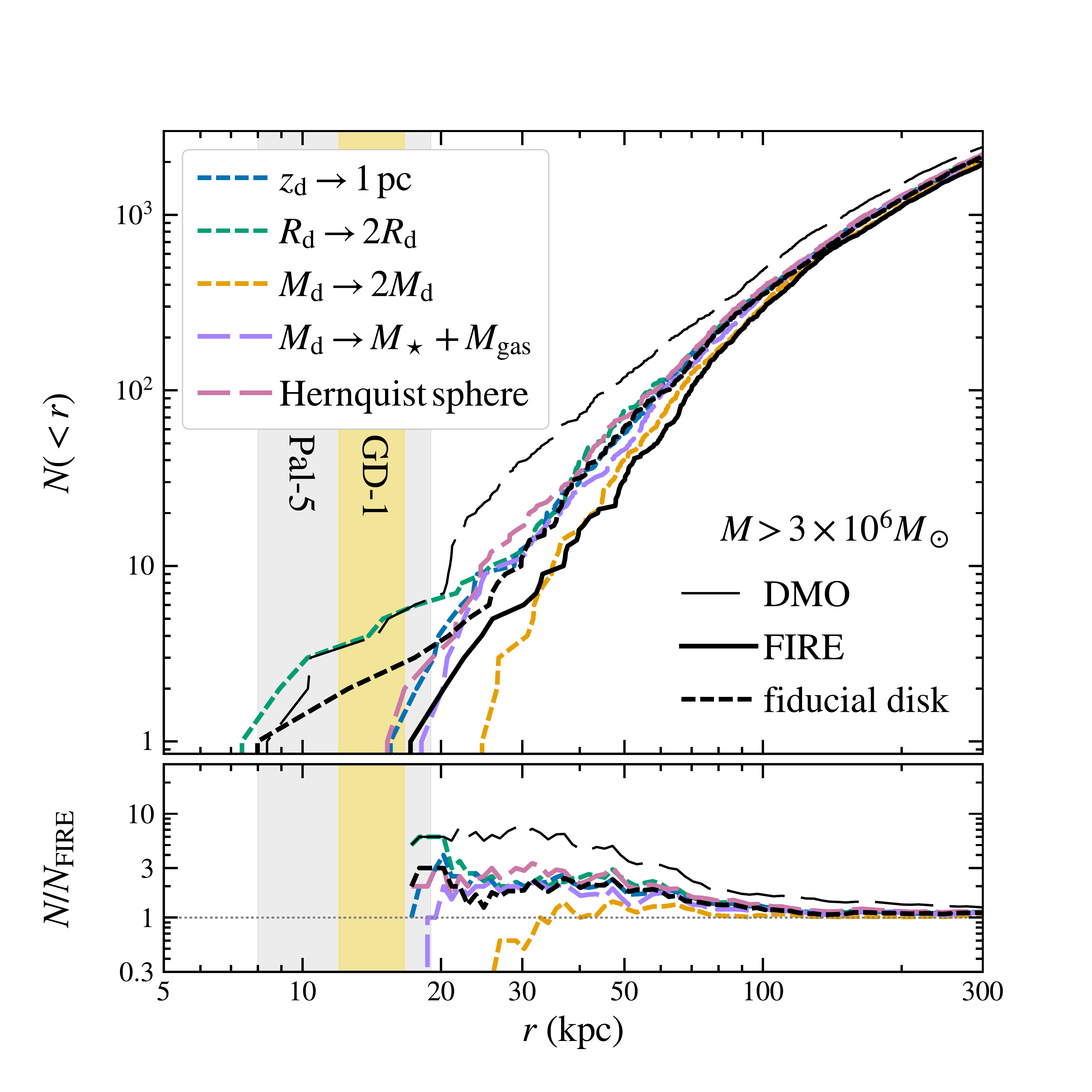}
\caption{
    How varying the properties of the embedded disk in \mi\ affects the subhalo population at
    $z = 0$. Black lines are identical to Figure~\ref{fig:vmaxdist}.  The colored dashed
    lines show embedded disk models with factors of 2 changes the scale radius, $\Rd$ and the
    total mass $\Md$, with a 1 pc thin disk, and with a total mass $\Md$ equal to the stellar
    mass plus the gas mass within the galaxy, while the dashed light red line shows the results
    of adding the potential from a Hernquist sphere of identical total mass to the simulation;
    details are provided in the text.
    \emph{Left:} Cumulative counts of subhalos above a given bound mass, $M$, within $100~\kpc$
    of the center at $z = 0$.
    \emph{Right:} Cumulative counts of resolved subhalos within a given radius, $r$.
    The results are largely independent of the radius or thickness of the disk, indicating that
    the shape and orientation of the potential are sub-dominant factors -- the spherically
    symmetric \citet{Hernquist1990} potential destroys nearly as many subhalos as the disk,
    though the excess in this run relative to the fiducial disk suggests that disk shocking
    is responsible for some of the destruction.   However, doubling the mass of the disk (orange
    dashed line) has a larger effect and causes significantly more subhalo destruction, leading
    to somewhat better agreement with the FIRE baryonic simulation, particularly within $\sim50~\kpc$.
    This improved agreement is likely because we originally match to only the stellar component
    of the central disk; doubling the disk mass roughly accounts for non-negligible gaseous contribution,
    particularly at $z \gtrsim 1$.  Overall, we emphasize that even imperfect fits to the disk provide
    a more accurate description of the subhalo population predicted by the fully baryonic simulation
    at nearly all masses and radii than a purely DMO realization.
}
\label{fig:compparams}
\end{figure*}

In order to explore the dependence on the specific disk potential that we add to the
dark matter simulations, we additionally simulate \mi\ with modified disk parameters.
Specifically, we hold two of the three disk parameters, $\Rd(z)$, $\zd(z)$, and $\Md(z)$
(see Equation~\ref{eqn:phi}) at their fiducial values, while varying the third. We
investigate: $\Rd(z) \rightarrow 2 \Rd(z)$, spreading the same amount of mass over a
larger surface area; $\zd \rightarrow 1~\pc$ at all times, forcing a nearly infinitesimally
thin disk; and $\Md(z) \rightarrow 2 \Md(z)$, making the disk more massive.  We also test
the dynamical importance of the gas in the galaxy by holding $\Rd$ and $\zd$ fixed at their 
fiducial values, but setting $\Md$ equal to the fiducial (stellar) mass plus the mass in
gas within $r_{90}$ (i.e., $\Md \rightarrow M_\star + M_{\rm gas})$.  In order to further test 
the importance of the shape and orientation  of the galactic potential, we also simulate \mi\ 
with a \citet{Hernquist1990} sphere, rather than the Miyamoto-Nagai disk potential.

Figure~\ref{fig:compparams} show the bound mass functions (left) and cumulative radial
distributions (right) of these simulations, along with the three versions of $\mi$
presented previously. The solid, dashed, and dotted black lines are identical to
those in the left panels of Figure~\ref{fig:vmaxdist}, while the green, blue, and
orange lines indicate the results of the simulations with $2\Rd$, $\zd=1~\pc$, and
$2\Md$, respectively.  The lavender lines plot counts obtained by additionally 
including the gas in the mass in the disk, and the light red dashed line plots subhalo 
counts with a Hernquist potential of identical mass and radial extent, quantified by 
$r_{90}$.

In summary, subhalo depletion most directly correlates with the mass of the central
disk.  The simulation with the $2\Md$ disk results in fewer subhalos, especially at
$r \lesssim 70 \,\kpc$.  The $2\Rd$ disk yields slightly more subhalos at small $r$,
while the dependence on $\zd$ is weak, with the $\zd=1~\pc$ disk predicting slightly
fewer subhalos than the fiducial disk.  The shape, and consequently the orientation,
of the potential appears to be of secondary importance, with the $\zd = 1~\pc$ disk
and the Hernquist sphere both yielding very similar results to the fiducial disk.
However, the slight excess in the simulation with a Hernquist potential, relative
to the fiducial disk, suggests that some subhalos are destroyed by disk shocking that
would otherwise survive the enhanced tidal forces \citep[also see][]{DOnghia2010}.
Therefore, small gains may be achieved by matching the orientation of the real
galaxy, but we emphasize again that even a disk with an imperfect, fixed orientation
yields subhalo populations with distributions that typically agree to within $\sim25\%$,
without the additional complexity of determining the correct orientation.

Interestingly, at most masses and radii, the $2\Md$ disk in fact yields a slightly
better match to the FIRE baryonic simulation than the fiducial disk simulation.\footnote{
The $2\Md$ simulation also provides a better match to the $\dperi$, $\vtan$, and $\vrad$
distributions of the FIRE baryonic simulations than the fiducial disk.}
As noted in Section~\ref{ssec:hydrosims}, we match the mass of the fiducial disk to the
stellar disk that forms in the baryonic simulation, but this neglects the (complex) contribution
from (fluctuating) gas in/near the disk.  Therefore, we \emph{may} posit that this 
improved match is a consequence of more accurately matching the \emph{total} baryonic mass 
at the center of the halo at early times, when the disk is dominated by gas ($z \gtrsim 1$).

However, the simulation with $\Md \rightarrow \mstar + M_\mathrm{gas}$ differs only 
slightly  from the results of the fiducial disk:  adding the gas to the disk mass yields marginally 
fewer subhalos at fixed mass and radius.  Because the differences in the disk mass between this 
run and the fiducial run occur primarily at early times,\footnote{While adding the gas mass
drastically increases the mass of the disk at early times (by a factor of $\sim$10 at $z = 3$), 
the contribution after $z = 1$ is less than a factor of two, and the total mass is within 
$\sim15\%$ of the fiducial (stellar) mass by $z = 0.5$.} the close agreement between the
two runs suggests that the mass at early times is relatively unimportant, likely because the 
majority of subhalos fall into the MW-size host halo after $z = 1$ \citep[e.g.][]{Wetzel2015a}.
The lingering discrepancy between the $\mstar + M_\mathrm{gas}$ run and the fully 
hydrodynamical simulation further indicates that the gas does have important, if 
second-order, dynamical effects on subhalos due to a combination of internal 
feedback, ram pressure and viscous stripping in the hot halo, and interactions with
inflows and outflows.

Overall, we emphasize that factors of $\sim2$ changes to the parameters of the embedded
disks have a relatively modest effect on the subhalo population at $z=0$, as compared
with the much stronger difference from \textit{not} using an embedded disk. Moreover, our
experiments with the shape of the potential show that the detailed geometry and orientation
of the disk (difficult to predict in non-fully baryonic simulations) are of small, secondary
importance: what matters (to leading order) is a central potential of the correct baryonic
mass and radial size.  Thus, even if the exact galaxy that \emph{would} form in a given halo
is not completely constrained, one still can improve upon the predictions of DMO simulation
by inserting a central galaxy with parameters drawn from empirical scaling relations.

\subsection{Comparison with \citet{Sawala2016b}}

Recently, \citet[][hereafter S17]{Sawala2016b} demonstrated that the subhalo populations of four of
the hosts in the APOSTLE project \citep{Sawala2016,Fattahi2016}, which simulates MW/M31-like
pairs of halos using the EAGLE prescriptions for star formation and feedback \citep{EAGLE},
are suppressed relative to their DMO counterparts.  Here we explicitly compare their results to
the differences that arises between the FIRE simulations and their DMO analogues.

Figure 1 of \citetalias{Sawala2016b} indicates that at the lowest masses resolved by the APOSTLE simulations, 
the DMO counterparts overpredict subhalo counts within 300 kpc by a factor of $\sim$1.3.  While the DMO 
analogues of the Latte simulations overpredict subhalo counts within 100 kpc by a factor of $\sim$2-3 
(Figure~\ref{fig:vmaxdist}), the ratio of $N(>M)_\mathrm{DMO}/N(>M)_{FIRE}$ within 300 kpc is roughly 
1.2 - 1.5 (Figure~\ref{fig:300kpc}), in close agreement with the results of \citetalias{Sawala2016b}.  
At smaller radii, however, Figure 3 of \citetalias{Sawala2016b} suggests that the DMO simulations 
typically overpredict counts within 50 kpc by a factor of $\sim$1.75 - 2, while our results 
(Figure~\ref{fig:50kpc}) suggest that DMO simulations overpredict counts within 50 kpc by a factor of 3 - 4.
Accordingly, we predict a larger discrepancy (between DMO and hydrodynamical simulations) in the radial 
distributions at small distances:  subhalo densities at $r \sim 30$~kpc are a factor of $\sim$4 times 
higher in the DMO analogues than in the FIRE simulations, while Figure~4 of \citetalias{Sawala2016b} 
indicates that their DMO analogues only predict a factor of $\sim$1.6 discrepancy at the same radius.

Therefore, while our general conclusions regarding depleted radial distributions agree well, results 
from the FIRE simulations suggest that the discrepancy grows larger at small radius ($r < 100$~kpc), but 
the results of \citetalias{Sawala2016b} indicate a roughly constant offset within $\sim$50 kpc.  While
this disagreement is second-order, it may be due to differences in the central masses of the galaxies
that form in the two sets of simulations.  Per Figure~7 of \citet{Fattahi2016}, the APOSTLE disks 
tend to lie slightly below expectations from abundance matching analyses.  Both \mi\ and \mf, 
however, fall slightly above those same relationships \citep{FIRE2}.  As evidenced by the parameter
survey presented in \S\ref{ssec:params}, increasing the mass density at the center of host destroys
more satellites at small radii without strongly altering results at $r \gtrsim 100~\kpc$. 

We caution, however, that the results in \citetalias{Sawala2016b} are based on the subhalo populations of 
only four hosts, and that our results include only two hosts.  Because some degree of halo-to-halo scatter 
is expected, it's not clear that our results should agree exactly with those of \citetalias{Sawala2016b}.  
It is encouraging, therefore, that all six hosts demonstrate a similar trend:  a clear reduction in subhalo 
populations, with the largest differences occuring within $\sim50$ - 100~kpc and with subhalos on radial 
orbits (low tangential velocities) preferentially depleted (Figure 8 of \citetalias{Sawala2016b}).

\section{Implications}
\label{sec:discussion}

Dark matter-only simulations have been used extensively to interpret
data and to make predictions for subhalo observables. Here we briefly
discuss several such investigations that are likely affected by our
results and speculate how the effects of a central galaxy potential in
more massive halos could affect the broader use of dark matter simulations
to interpret data.

As discussed in \S\ref{sec:intro} and \S\ref{sec:results}, the significant
depletion that we see in our baryonic and embedded disk simulations at small
radii has important implications for ongoing searches for substructure in the
MW halo. Perhaps the most promising method involves looking for gaps and other
anomalies in the stellar streams of Palomar-5 and GD-1. Both of these streams exist
within $20$ kpc of the Galaxy, therefore they sit within the region that is most
severely depleted (see Figure~\ref{fig:vmaxdist}). Our results imply that there
may be no dark substructures in such a region today to affect these streams, and
that even if they orbited in this region in the past, their lifetime is much
shorter than predicted from DMO simulations.  However, further work is needed to
sample full subhalo evolutionary histories, and as we discuss in the next section,
a more focused effort on simulating tuned MW analogues (set to the mass of the MW galaxy,
rather than being matched to a comparison hydrodynamical simulation) with a 
statistical sample of initial conditions will be required to determine the range of subhalo 
counts that we expect around our Galaxy within this region.

More generally, any analyses that depends strongly on the subhalo mass function,
the radial distribution, or the velocity distribution, particularly near the center
of the host, will be severely impacted by the presence of a central galaxy. One
example is the expected completeness corrections to the faint-end of the MW stellar
mass function \citep{Tollerud2008, Hargis2014}. This depends non-trivially on the
radial distribution of subhalos, because the correction is based on how many more
subhalos we expect to exist within some large volume (for example, $r \le 400$ kpc)
given an observed number within a smaller (for example, $r \le 100$ kpc), observationally
complete volume. The enhanced central depletion in our simulations suggests that there
are fractionally even more satellites waiting to be discovered at large radii, given
the number that we currently observe within the completeness limits of SDSS, DES, or
Pan-STARRS.  As pointed out by \citet{Ahmed2016}, who also found that subhalo
populations are more radially extended in baryonic simulations than in DMO, depleting
the central region of substructure has the additional effect of increasing the
statistical significance of any potential ``planes of satellites.''

The missing satellites problem is clearly reduced in severity by the destruction
of subhalos with $M \sim10^9$--$10^{10} \Msun$. Similarly, the tension in comparing the
number of dwarf galaxies containing old stars around the MW with expectations from the observed steep
luminosity functions during the reionization era, as discussed by \citet{BoylanKolchin2014},
will be reduced; many of those ancient stars instead would be dispersed into the stellar
halo via the enhanced disruption brought on by the central disk. Similarly, the constraints
on stellar mass functions at $z \sim5$ based on MW galaxy counts presented in
\citet{Graus2016} will be modified.

Limits on warm dark matter models \citep[e.g.][]{Horiuchi2014, Horiuchi2016} associated
with requiring enough subhalos to host the known satellites will become even tighter.
Predictions for the substructure boost for dark matter annihilation signals in the Galactic
center \citep{Kuhlen09, Bovy2009} will decrease. However, subhalos would still be expected to
contribute to a boost at larger radii, which could be important for indirect detection searches
around the M31 halo or in the all-sky background from the MW halo itself.

The reduction in subhalo counts also modifies the results of using counts of satellite galaxies
in the Local Group to constrain the $\mstar - \mhalo$ relation at low masses
\citep[e.g.][]{ELVIS, Brook2014, GK2016}. Specifically, matching to fewer dark matter
halos, as implied by our results, requires a flatter log-slope, which also shifts the
relation closer towards one that alleviates the `too-big-to-fail' problem
(\citealp{BK2011,BK2012,ELVISTBTF,Papastergis2015}, and also see \citealp{Jethwa2016}).
The effect of enhanced disruption from a central disk in alleviating the too-big-to-fail
problem was first emphasized by \citet{BrooksZolotov2012}.

The preferential destruction of subhalos on plunging orbits also has consequences
for analyses that utilize the velocity distributions of subhalos in DMO simulations.
For example, the number of `backsplash galaxies'
\citep[galaxies beyond the virial radius of a more massive host that orbited within the virial radius in the past; e.g.][]{Teyssier2012,Wetzel2014,ELVIS}
that have passed close to the central galaxy should be drastically reduced, though the number
that have had only glancing encounters with the host halo will be largely unaltered. Similarly,
the works of \citet{Fillingham2015} and \citet{Wetzel2015b} explored the timescales over which
the satellite galaxies of the MW and M31 had their star formation quenched after infall using subhalo
catalogs taken from the ELVIS simulations. While they found that the environmental quenching
timescales of satellite dwarf galaxies are short ($\lesssim 2$~Gyr; much shorter than for more
massive satellites), our results imply \textit{even shorter} quenching timescales, because many
of the subhalos that fell in at early times have preferentially smaller pericenters \citep{Wetzel2015a}
and therefore should be destroyed by the central galaxy. Because satellites with low tangential
velocities also are destroyed with high efficiency, however, the constraints on ram-pressure
stripping from \citet{Fillingham2016} may become less stringent.  Similarly, the recent
discrepancy pointed out by \citet{Cautun2016}, wherein the MW satellites are on more
tangentially biased orbits than expected from \lcdm, is alleviated by the destruction of
these low $\vtan$ subhalos.

While our analysis focused on MW-size halos, substructure depletion caused by a central galaxy is
likely important around larger ($\mhalo \sim10^{13} \Msun$) galaxies as well. This is particularly
relevant for searches for dark subhalos via lensing anomalies
\citep[e.g.][]{Vegetti2010,MacLeod2013,Nierenberg2014,Hezaveh2016}.
However, the strength of the impact of these more massive galaxies on their subhalos is unclear. The
MW represents the mass at which the ratio of central galaxy stellar mass to host halo mass is highest
\citep{Leauthaud2012,Behroozi2013,Moster2013}, so this is the mass scale at which the central galaxy
most strongly affects the halo potential.  Thus, the effect in larger hosts may be weaker. Furthermore,
more massive halos generally assemble at later times (with satellites falling in later), suggesting that
subhalo destruction may be further suppressed relative to our findings here.

In fact, \citet{Fiacconi2016} recently presented two baryonic simulations of galaxies in this
mass regime, and found that baryonic contraction actually increases the number of massive subhalos
near the center of the host relative to DMO simulations. However, Graus et al. (in preparation)
perform a similar analysis using the Illustris and Illustris Dark simulations \citep{Illustris1,Illustris2}
and find the opposite effect: the baryonic simulations have fewer subhalos at $\sim10^8 \Msun$ around
lens hosts, with the largest deficit near the central regions. \citet{Despali2016} similarly compared
the DMO and baryonic versions of the Illustris and EAGLE \citep{EAGLE} simulations, and demonstrated
that subhalo counts around hosts of mass $10^{12.5}$--$10^{14} \Msun$ are suppressed by $\sim30\%$
at subhalo masses $10^9 \Msun$ in the baryonic simulations, with the largest suppression near
the center of the hosts (also see \citealp{Chua2016}, who reached similar conclusions using the Illustris
simulations).  The impact on the lowest-mass subhalos, however, remains largely unexplored due to the
difficulty of simulating galaxies of this mass at high resolution with baryonic physics.

Finally, if a similar degree of substructure depletion occurs for roughly LMC to MW-size
subhalos ($\mhalo \sim10^{11}$--$10^{12} \Msun$) around massive hosts
($\mhalo \gtrsim 10^{14} \Msun$), it could have important implications for the use of DMO
simulations to interpret small-scale (`one-halo') clustering statistics through subhalo
abundance matching \citep[e.g.][]{Conroy2006, Reddick2013}, as compared with studies that
examined these trends for subhalo disruption without modeling a central galaxy
\citep[e.g.][]{WetzelWhite2010}. Specifically, if the subhalos that are massive enough to
host bright galaxies ($\mstar \gtrsim 10^{10} \Msun$) are preferentially depleted in galaxy
groups owing to the potential of the brightest cluster galaxy, then abundance matching
analyses would tend to assign too little stellar mass to subhalos of a given $M$ to avoid
overproducing those galaxies. If this effect is important in groups and clusters, then it
may also influence the use of similar approaches to understand trends between galaxy color
and age with subhalo accretion times \citep[e.g.][]{Mahajan2011,Wetzel2013,Hirschmann2014,Oman2016}.
However, the increased concentrations of MW-mass subhalos in baryonic simulations may also 
make these systems more resistant to stripping \citep[e.g.][]{Chua2016}, such that even if 
there are fewer satellites overall, a greater proportion of those remaining would have 
$\mhalo\sim10^{12}\msun$ than DMO simulations would predict.

\section{Conclusions}
\label{sec:conclusions}

The interplay between dark matter and baryons, and the subsequent importance of
baryons in correctly predicting the properties of DM halos with cosmological
simulations has a rich history in this field \citep[e.g.][]{diCintio2011,Governato2012,
Pontzen2012,Zolotov2012,BrooksZolotov2012,Arraki2014,bentbybaryons,diCintio2014,
Onorbe2015,Chan2015,Sawala2016,Cui2016,Wetzel2016}. In this paper, we explored
the role of baryons in affecting the low-mass subhalos of interest for observational
searches for dark subhalos: dark matter mass $M > 3 \times 10^6 \Msun; \: V_{\rm max} > 5~\kms$.

Our exploration relied on two MW-mass dark matter halos from the Latte simulation
suite \citep{Wetzel2016}, each simulated (1) with full baryonic physics from the
FIRE project, (2) with dark matter only, and (3) with dark matter plus an embedded
disk potential for the central galaxy that evolves to match the stellar disk in the
corresponding baryonic simulation. As shown in Figure~\ref{fig:vmaxdist}, relative
to the DMO simulations, subhalo counts in the baryonic simulations are lower by a
factor of $\sim5$ within $25$ kpc of the halo center. Both of the baryonic simulations
are \textit{completely} devoid of substructure within $15$ kpc at $z = 0$, though
$\sim20$ subhalos with pericentric distances $<20~\kpc$ survive to the present day.  
Subhalo depletion becomes less important with increasing radius: within $\sim300$ kpc (roughly
the halo virial radius), subhalo counts in the baryonic simulations are lower by only
$\sim15$--$30\%$ compared to the DMO simulations.  This depletion is driven by
the preferential destruction of subhalos on radial orbits that get to
$d_{\rm peri} \lesssim 30 \,\kpc$ from the halo center. This in turn biases the orbital
velocities of \textit{surviving} subhalos at $z = 0$ to be more tangential in simulations
that have central galaxies than they are in DMO simulations.

Importantly, the simulations with embedded disk potentials reproduce at least $\sim75\%$
of the depletion evident in the mass functions and radial distributions. This
good agreement is important for two reasons.  First, because we carefully matched the
evolving disk potential to the stellar disk in the baryonic simulations, this comparison
provides a clear physical explanation for the origin of subhalo depletion in the baryonic
simulations: the bulk of the subhalo depletion arises simply from the tidal field of the
central galaxy itself.  These results also imply that the majority of subhalo depletion
is \textit{independent} of the exact model of feedback or star formation, because our
embedded disk simulations include none of these.  However, we emphasize that we matched
the evolving disk potential to the {\em stellar mass} in the baryonic simulations; this
therefore provides a conservative under-estimate of the total baryonic mass of the central
galaxy, especially at early times when the simulated galaxies were gas rich. As
Figure~\ref{fig:compparams} showed, though, adding the approximate gas mass to the disk 
yields only marginally better agreement between the embedded disk and baryonic simulations.
Therefore, a small fraction of the substructure depletion is \emph{not} 
due to interactions with the central galaxy, and cannot be captured with this technique.

Second, given its significantly improved accuracy as compared with DMO simulations, our
method of embedding a central disk potential provides an inexpensive way to significantly
improve the accuracy of DMO simulations.  The population statistics of surviving subhalos
in the embedded disk simulations display much better agreement with those in the FIRE
baryonic simulations in every statistic that we have checked, including infall times,
pericentric distances, radial velocities, total orbital velocities, radial profiles,
and counts as a function of mass at $z = 0$ or at infall.  We also find that the disk
simulations yield more accurate counts around the hosts at higher redshift, though the
differences are less dramatic than at $z = 0$.  Thus, one can vastly improve upon predictions
from purely DMO simulations, at nearly the same CPU cost, by simply including a central
galactic potential.

Moreover, as Figure~\ref{fig:compparams} showed, subhalo depletion is largely insensitive
to the central disk thickness, and it depends only mildly on disk radius and the detailed
shape of the potential.  The most important parameter is simply the central galaxy mass.
Thus, one can (to reasonable approximation) adopt a simple, spherically-symmetric analytic
potential with parameters taken from observations (e.g. abundance-matching) or large-volume
simulations.

The differences that we see between simulations with and without central galaxies are particularly
important for dark substructure searches that rely on cold stellar streams within 20 kpc of the
Galaxy.
However, it is difficult to estimate the expected halo-to-halo variance based
on just two halos. As shown in Figure~\ref{fig:dperi}, the intrinsic pericenter distribution
for a given subhalo population can vary considerably, which likely affects subhalo disruption
considerably. A larger suite of hosts simulated with embedded disks that match the MW is
required to make more concrete statistical statements (Kelley et al., in preparation).

We demonstrated that baryonic effects are crucially important for interpreting
ongoing substructure searches. Galaxies exist within the centers of halos and are dynamically
important within the vicinity of disks like the MW. Our method of embedding galactic
potentials in cosmological zoom-in simulations provides an avenue to producing more accurate
substructure predictions, relative to baryonic simulations, without the millions of CPU hours
required by those simulations. These embedded galaxies therefore hold the promise of improved
predictions for the statistical properties of subhalo populations, which are necessary to fully
interpret the results of many upcoming observations, including those aimed at detecting tiny, dark subhalos.

\section*{Acknowledgements}

The authors thank Jo Bovy, Mia Bovill, and Brandon Bozek for valuable discussions
and the anonymous referee for several suggestions that have improved the manuscript.  
The authors also thank Erik Tollerud for providing software used to create visualizations 
and Alexander Knebe, Peter Behroozi, and Oliver Hahn, respectively, for making 
\texttt{AHF}, \texttt{consistent-trees}, and \texttt{MUSIC} publicly available.

Support for SGK was provided by NASA through Einstein Postdoctoral Fellowship
grant number PF5-160136 awarded by the Chandra X-ray Center, which is operated
by the Smithsonian Astrophysical Observatory for NASA under contract NAS8-03060.
AW was supported by a Caltech-Carnegie Fellowship, in part through the Moore Center 
for Theoretical Cosmology and Physics at Caltech, and by NASA through grant 
HST-GO-14734 from STScI. JSB, and TK were supported by NSF grant AST-1518291
and by NASA through HST theory grants (programs AR-13921, AR-13888, and AR-14282.001) awarded
by the Space Telescope Science Institute (STScI), which is operated by the Association of
Universities for Research in Astronomy (AURA), Inc., under NASA contract NAS5-26555.
Support for PFH was provided by an Alfred P. Sloan Research Fellowship, NASA ATP Grant NNX14AH35G,
and NSF Collaborative Research Grant \#1411920 and CAREER grant \#1455342. MBK acknowledges support
from the National Science Foundation (grant AST-1517226) and from NASA through HST theory grants
(programs AR-12836, AR-13888, AR-13896, and AR-14282) awarded by STScI. CAFG was supported by
NSF through grants AST-1412836 and AST-1517491, and by NASA through grant NNX15AB22G.  DK
acknowledges support from NSF grant AST-1412153 and the Cottrell Scholar Award from the Research
Corporation for Science Advancement.  EQ was supported by NASA ATP grant 12-APT12-0183, a Simons
Investigator award from the Simons Foundation, and the David and Lucile Packard Foundation.
Support for ASG was provided by NSF grant AST-1009973.

Numerical calculations were run on the Caltech compute cluster `Zwicky' (NSF MRI award \#PHY-0960291)
and allocation TG-AST130039 granted by the Extreme Science and Engineering Discovery Environment (XSEDE)
supported by the NSF. Resources supporting this work were also provided by the NASA High-End Computing (HEC)
Program through the NASA Advanced Supercomputing (NAS) Division at Ames Research Center. This work also made
use of \texttt{Astropy}, a community-developed core Python package for Astronomy \citep{Astropy}, \texttt{matplotlib}
\citep{Matplotlib}, \texttt{numpy} \citep{numpy}, \texttt{scipy} \citep{scipy}, \texttt{ipython} \citep{ipython},
\texttt{Mayavi} \citep{mayavi}, and NASA's Astrophysics Data System.

\section*{Appendix A: Distributions within $50$ and $300~\kpc$}

\begin{figure*}
\centering
\includegraphics[width=0.5\columnwidth]{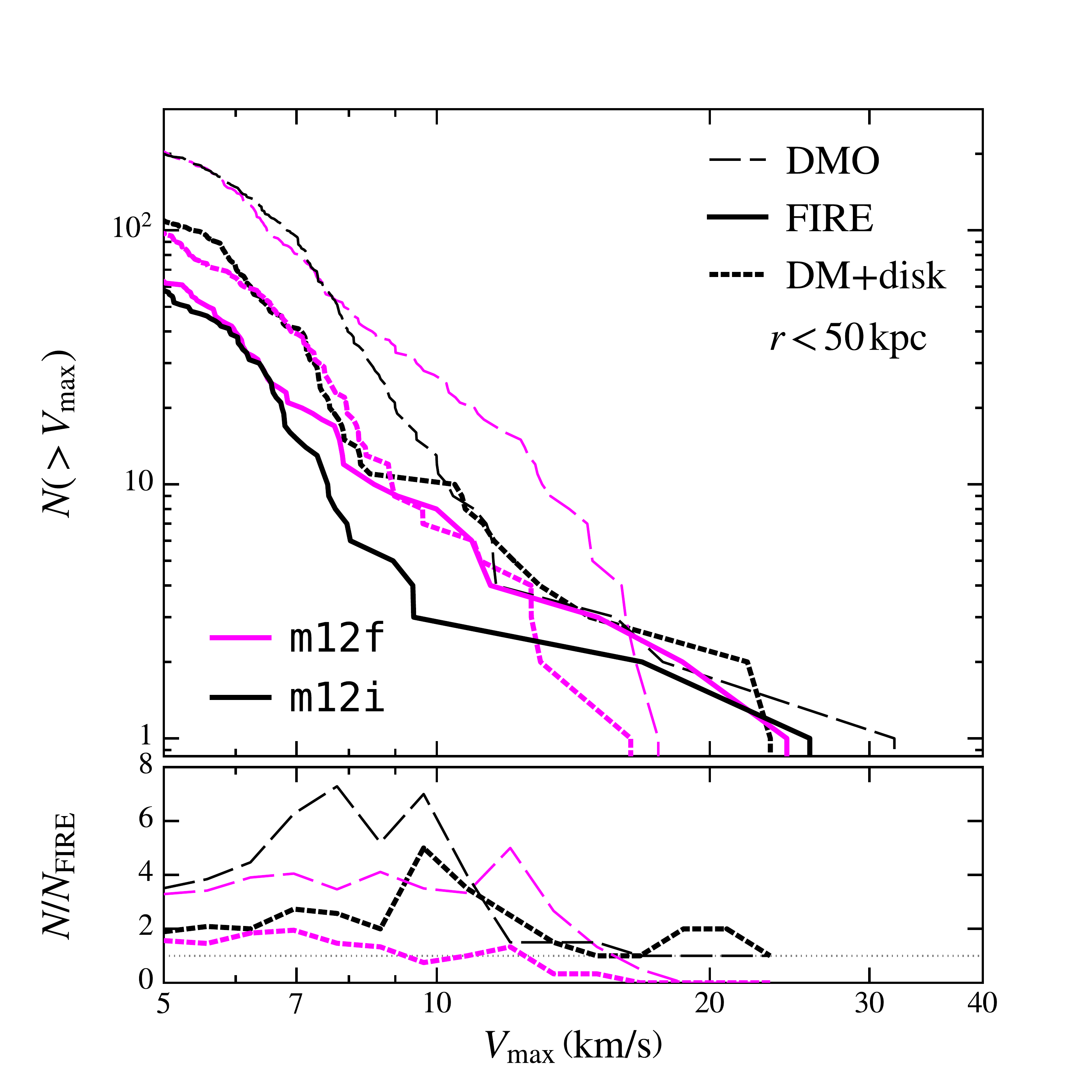}
\includegraphics[width=0.5\columnwidth]{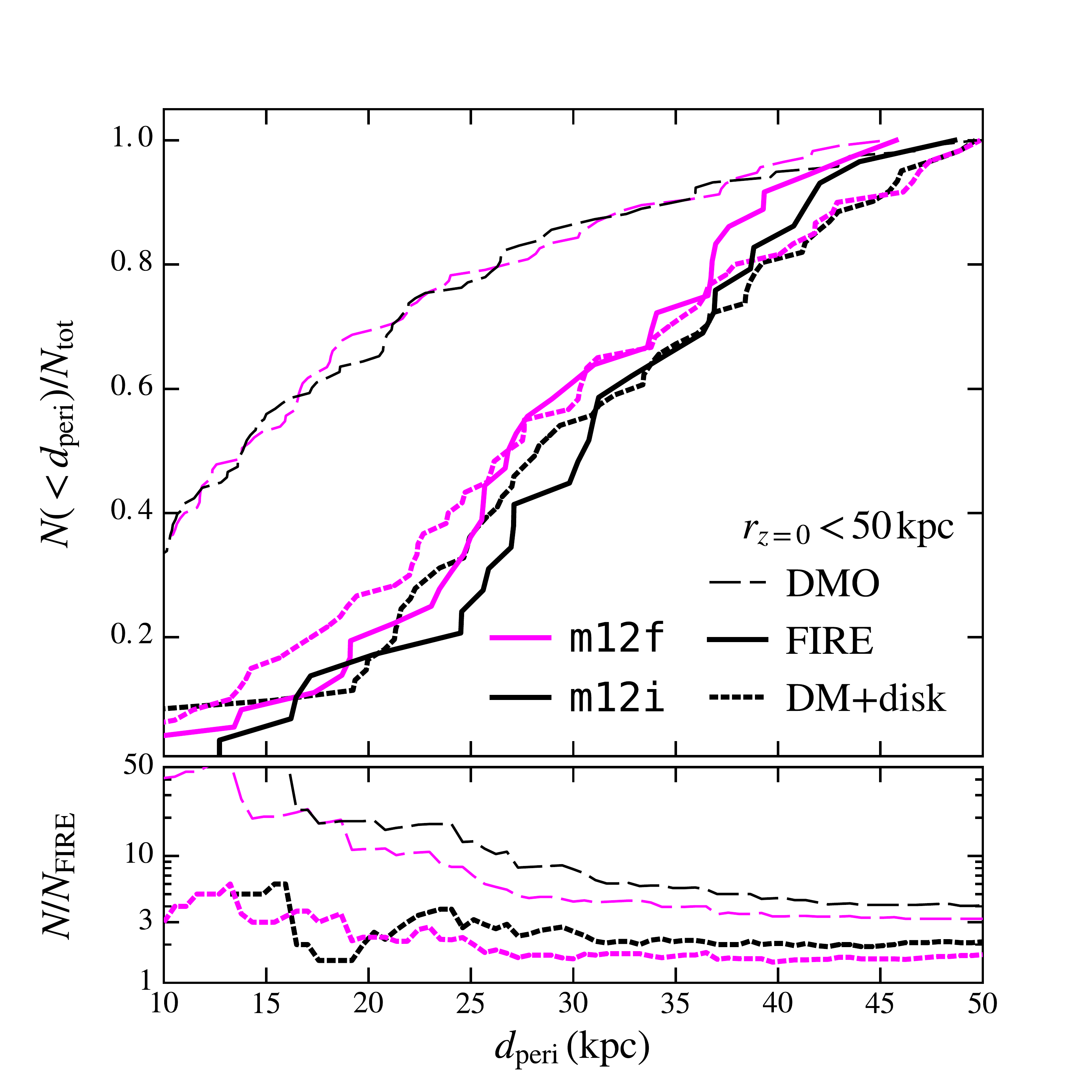} 
\includegraphics[width=0.5\columnwidth]{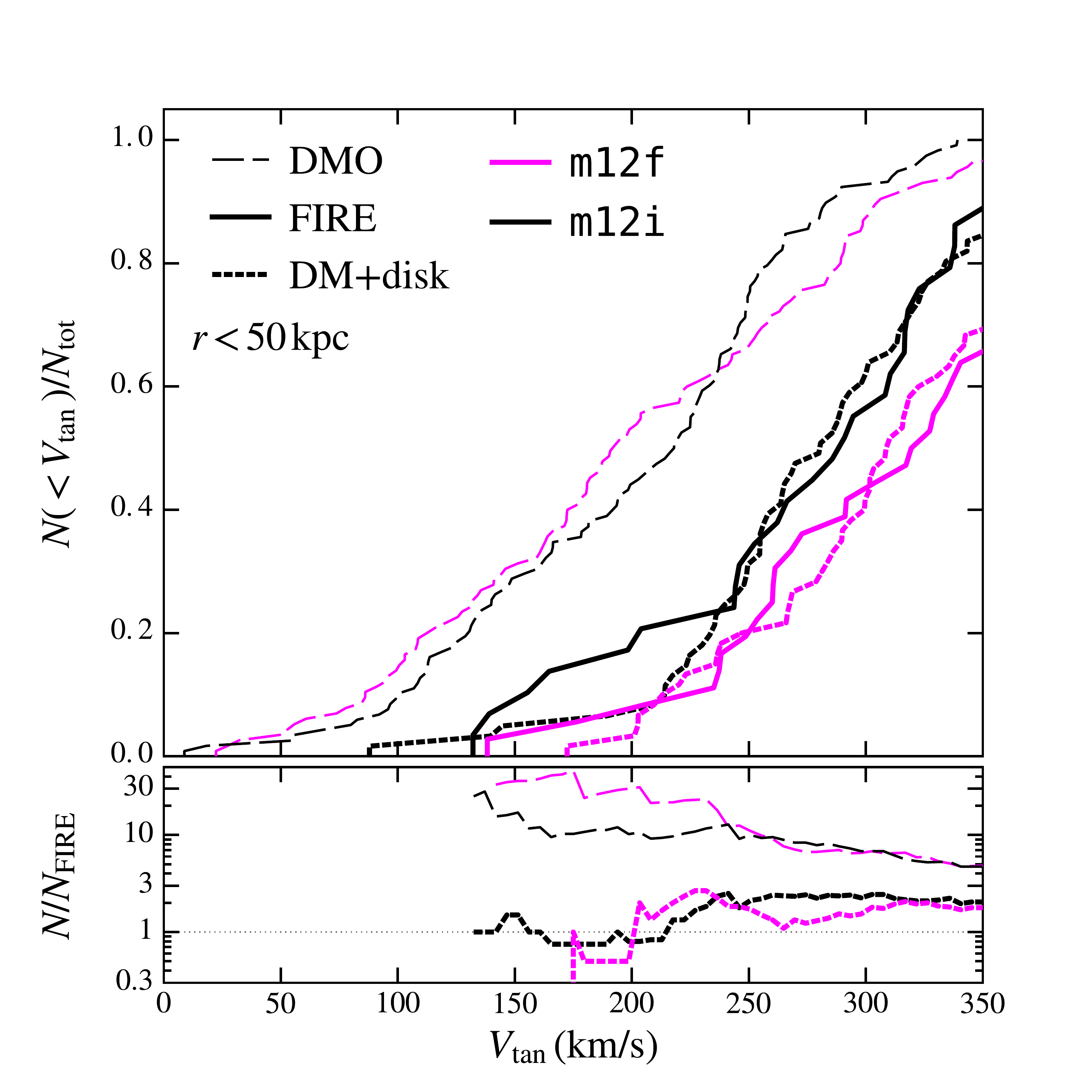}
\includegraphics[width=0.5\columnwidth]{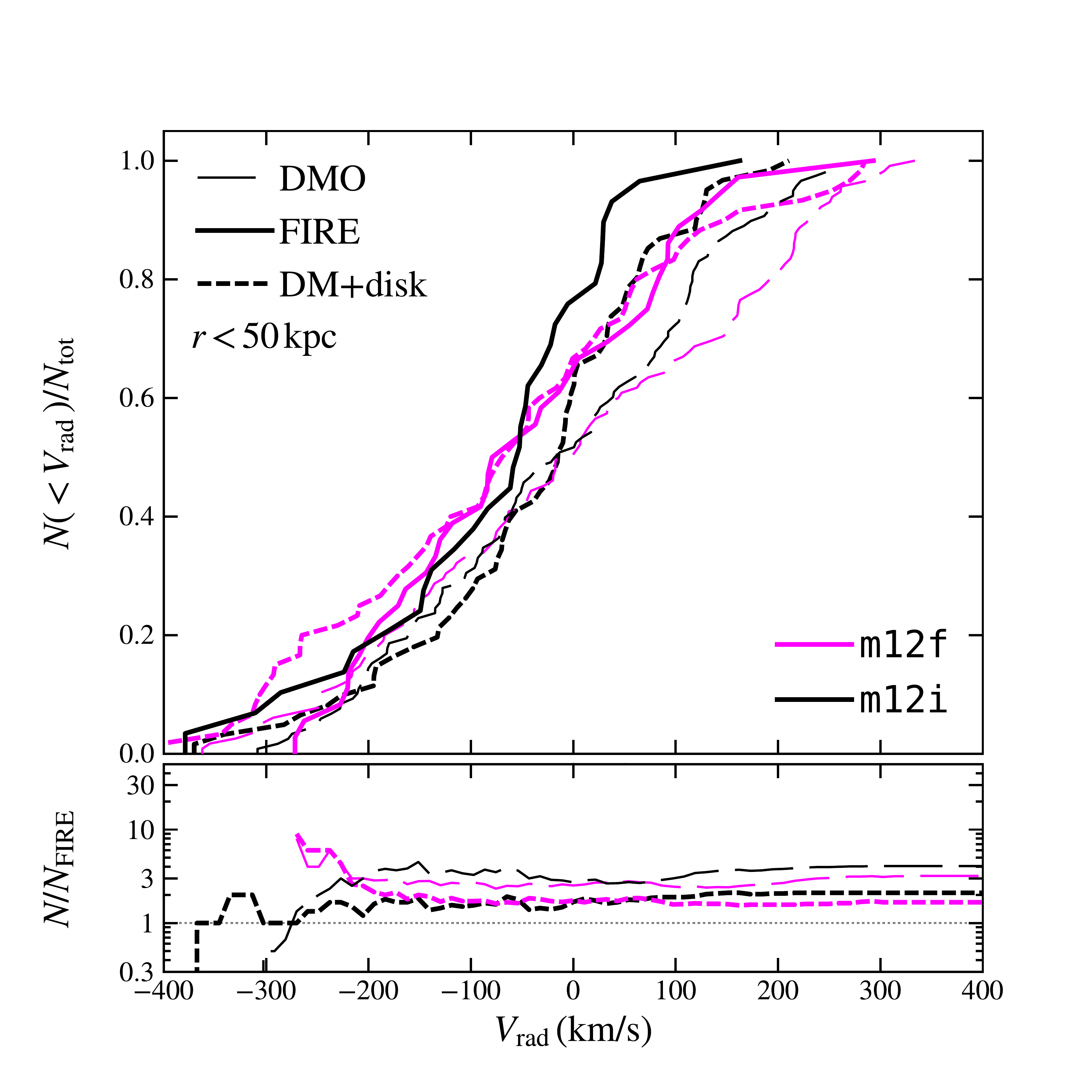}
\caption{
From \emph{left} to \emph{right}, cumulative counts of subhalo within $50~\kpc$, 
as a function of $\vmax$ (similar to the top panels of Figure~\ref{fig:vmaxdist}); 
normalized counts as a function of pericentric distance (similar to the left panel of
Figure~\ref{fig:dperi}); and normalized counts as a function of tangential and radial
velocity (similar to Figure~\ref{fig:velocities}). All trends are similar to those
presented in the main text: subhalos on radial orbits (low $\vtan$ and small
$\dperi$) are readily destroyed.  However, the overall destruction is higher within
$50~\kpc$: DMO simulations overpredict the total number of subhalos by a factor of
$3$--$4$, compared to factors of $1.5$--$2$ within $100~\kpc$.
}
\label{fig:50kpc}
\end{figure*}

\begin{figure*}
\centering
\includegraphics[width=0.5\columnwidth]{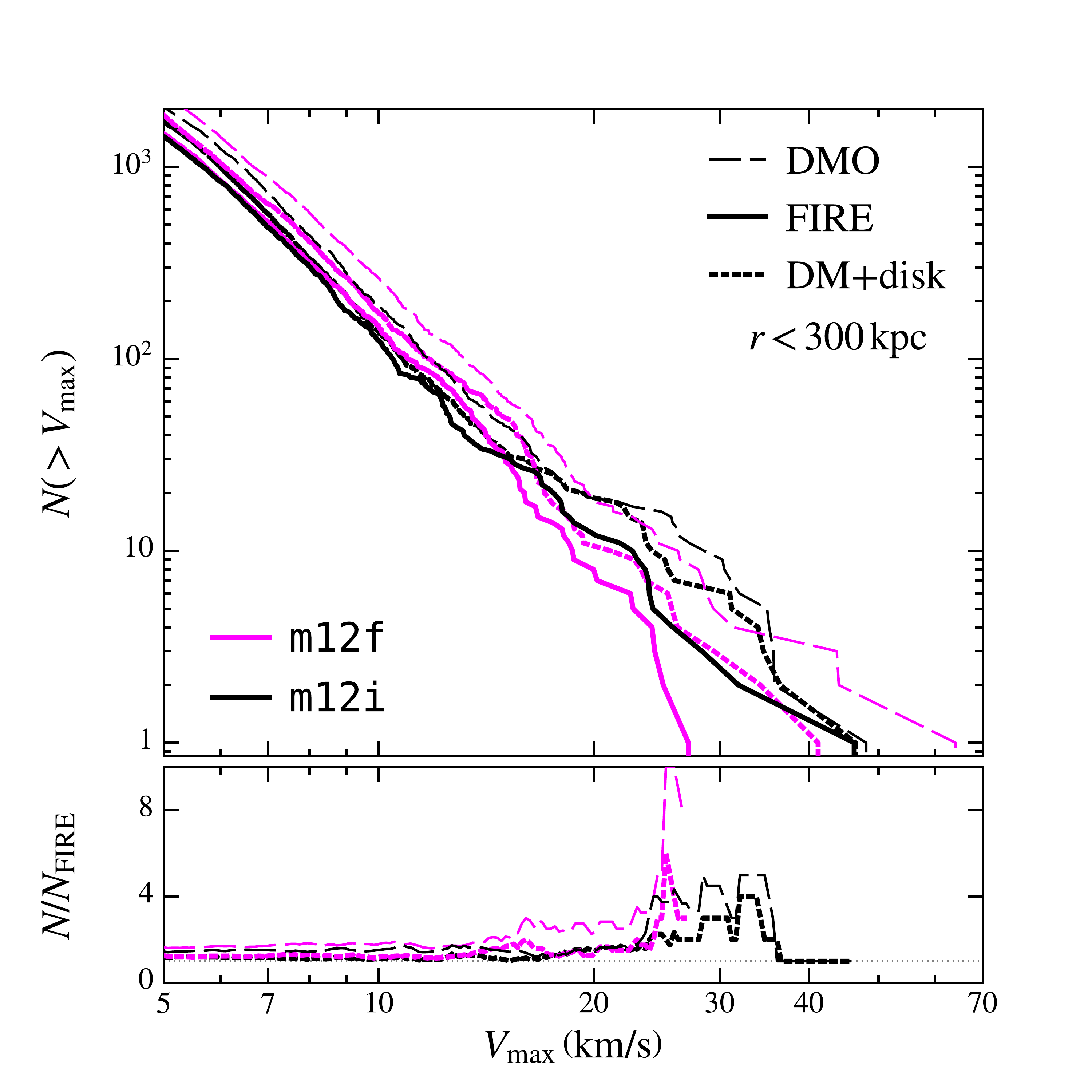}
\includegraphics[width=0.5\columnwidth]{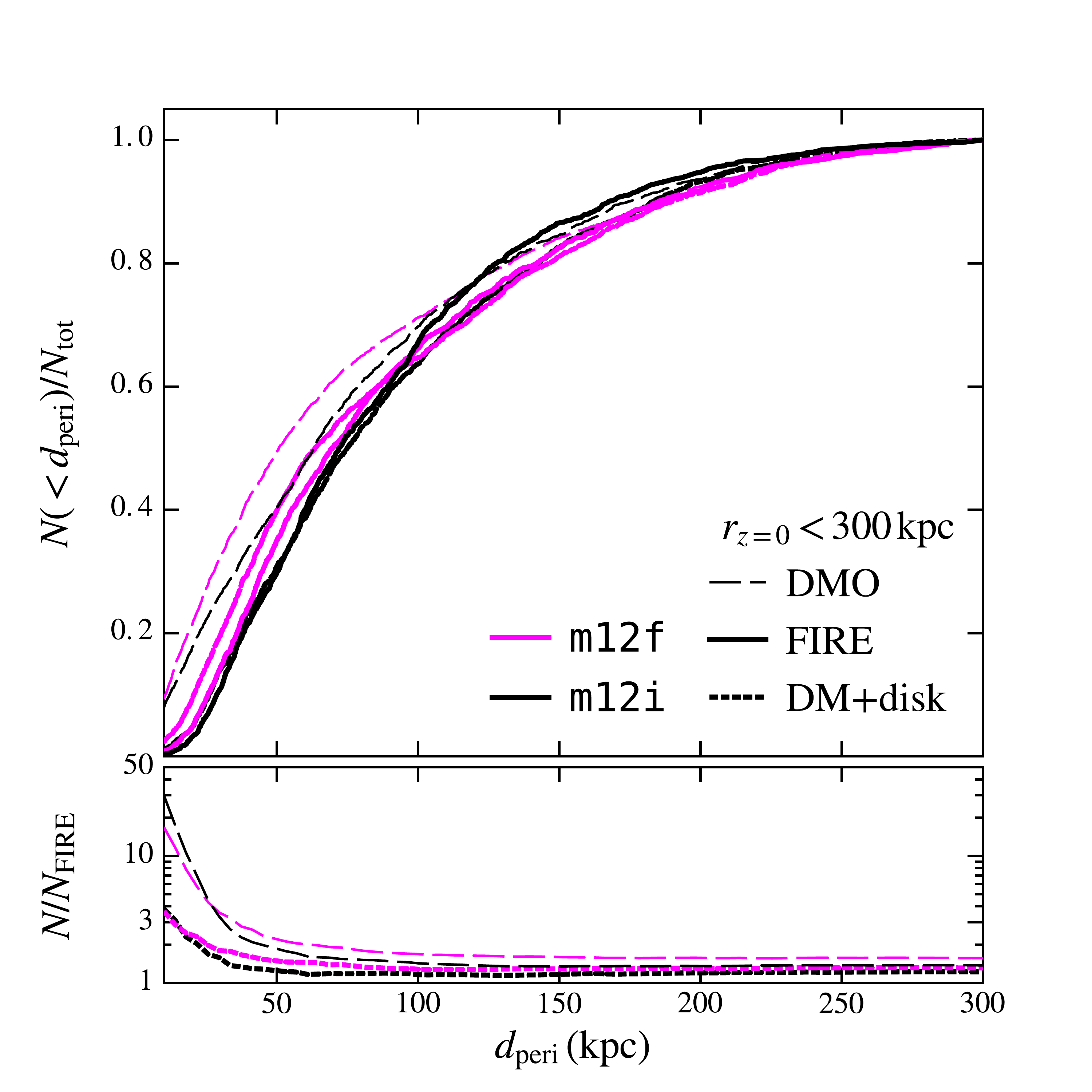} 
\includegraphics[width=0.5\columnwidth]{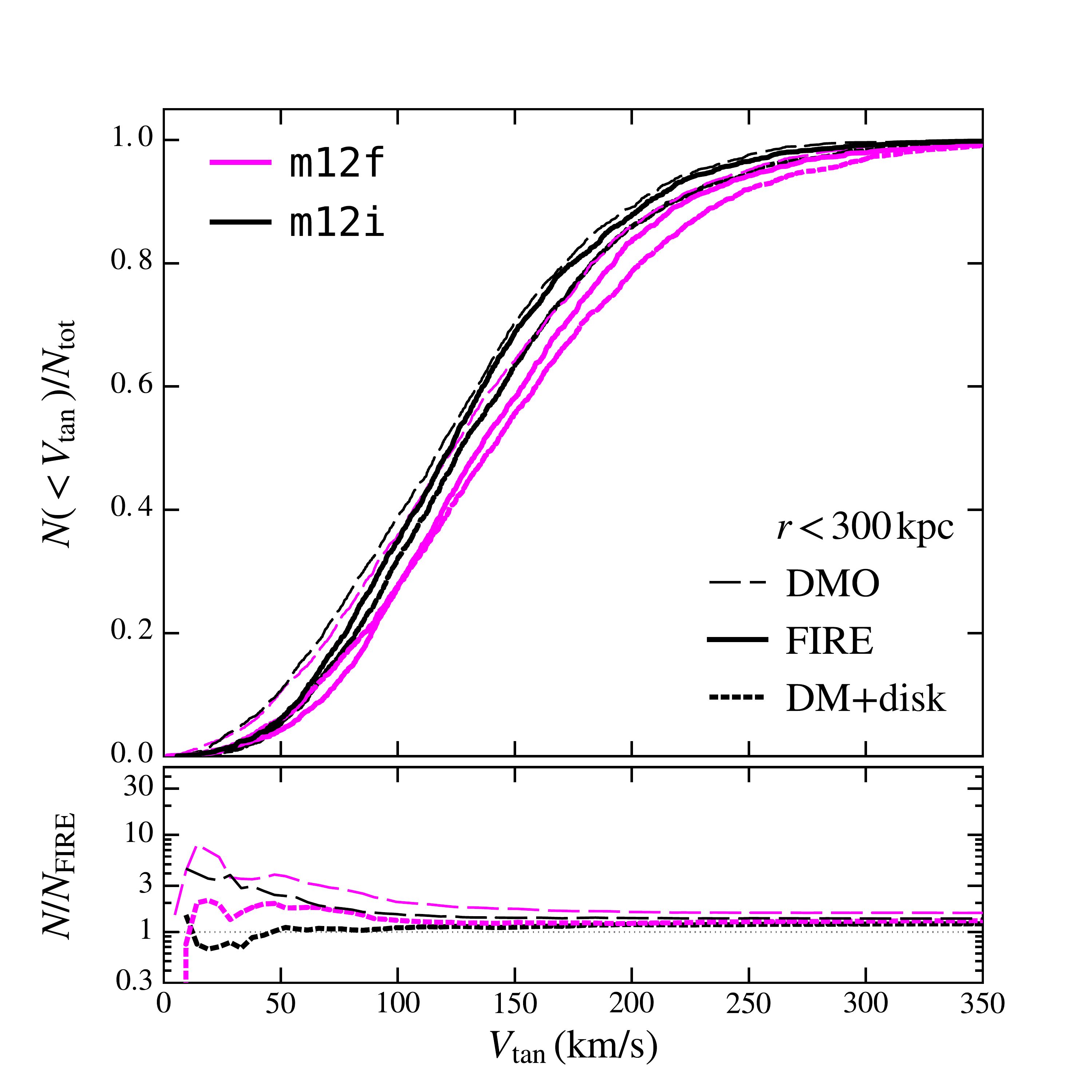}
\includegraphics[width=0.5\columnwidth]{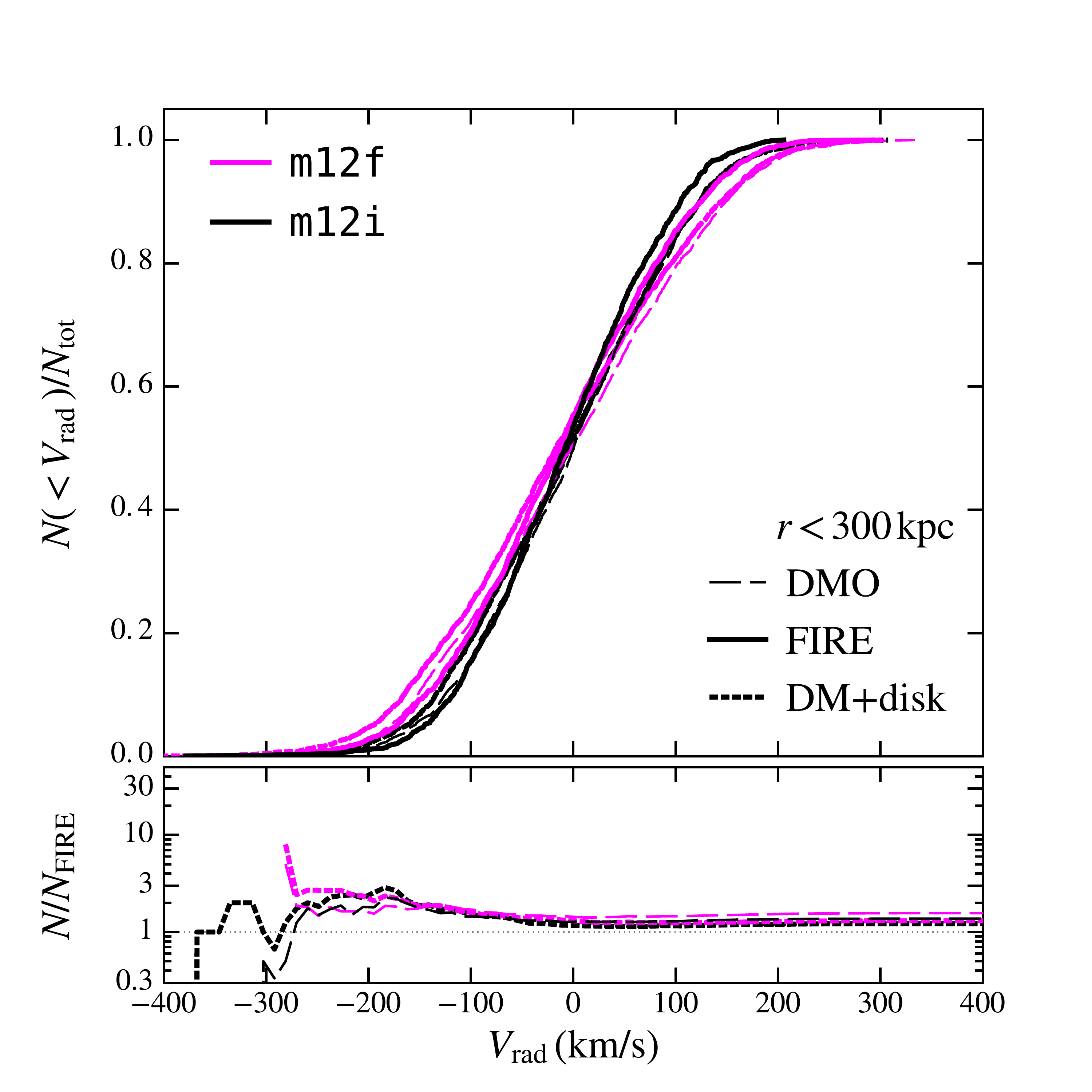}
\caption{
Identical to Figure~\ref{fig:50kpc}, but including all subhalos within $300~\kpc$,
roughly corresponding to the host halos' virial radii. The relative amount of destruction
is significantly lower than within $50$ or $100~\kpc$, but the embedded disk simulations
still yield a better match than the DMO simulations to the FIRE baryonic simulations.
Moreover, the imprint of the central galaxy remains on the $\dperi$ and $\vtan$ distributions,
even out to $300~\kpc$.
}
\label{fig:300kpc}
\end{figure*}

The Figures presented in the main body of the paper plot subhalo counts, as a
function of $\vmax$, $\dperi$, and $\vtan$, within $100~\kpc$. For completeness,
we plot here similar distributions, but including subhalos within either
$50~\kpc$ (Figure~\ref{fig:50kpc}) or $300~\kpc$ (Figure~\ref{fig:300kpc}) of the host centers. The fractional difference in the purely
DMO simulations are lower on $\sim300~\kpc$ scales than within either $50$ or
$100~\kpc$, but the impact of the central galaxy remains present in the
lack of subhalos with low $\vtan$ or small $\dperi$. Within $50~\kpc$, the
central galaxy is extremely destructive: only $\sim25\%$ of the subhalos present
in the DMO simulations survive in the presence of a disk. Similarly, no
subhalos with $\vtan \lesssim 100~\kms$ remain within $50~\kpc$ of the galaxy.

\section*{Appendix B: Resolution}

Because of the cost of simulating the systems at even higher resolution than
that presented in the main body, we instead establish that we reliably identify
substructures at the same particle count using lower-resolution simulations.
Specifically, we compare counts as a function of mass and radius to simulations
of \mi\ with particle masses $8$ times larger than our fiducial simulations
($m_{\rm p} = 3.4\times10^5\,\msun$) and with softening lengths a factor of two larger.
A detailed discussion of resolution in the FIRE simulations can be
found in \citet{FIRE2}.

\begin{figure*}
    \centering
    \includegraphics[width=\columnwidth]{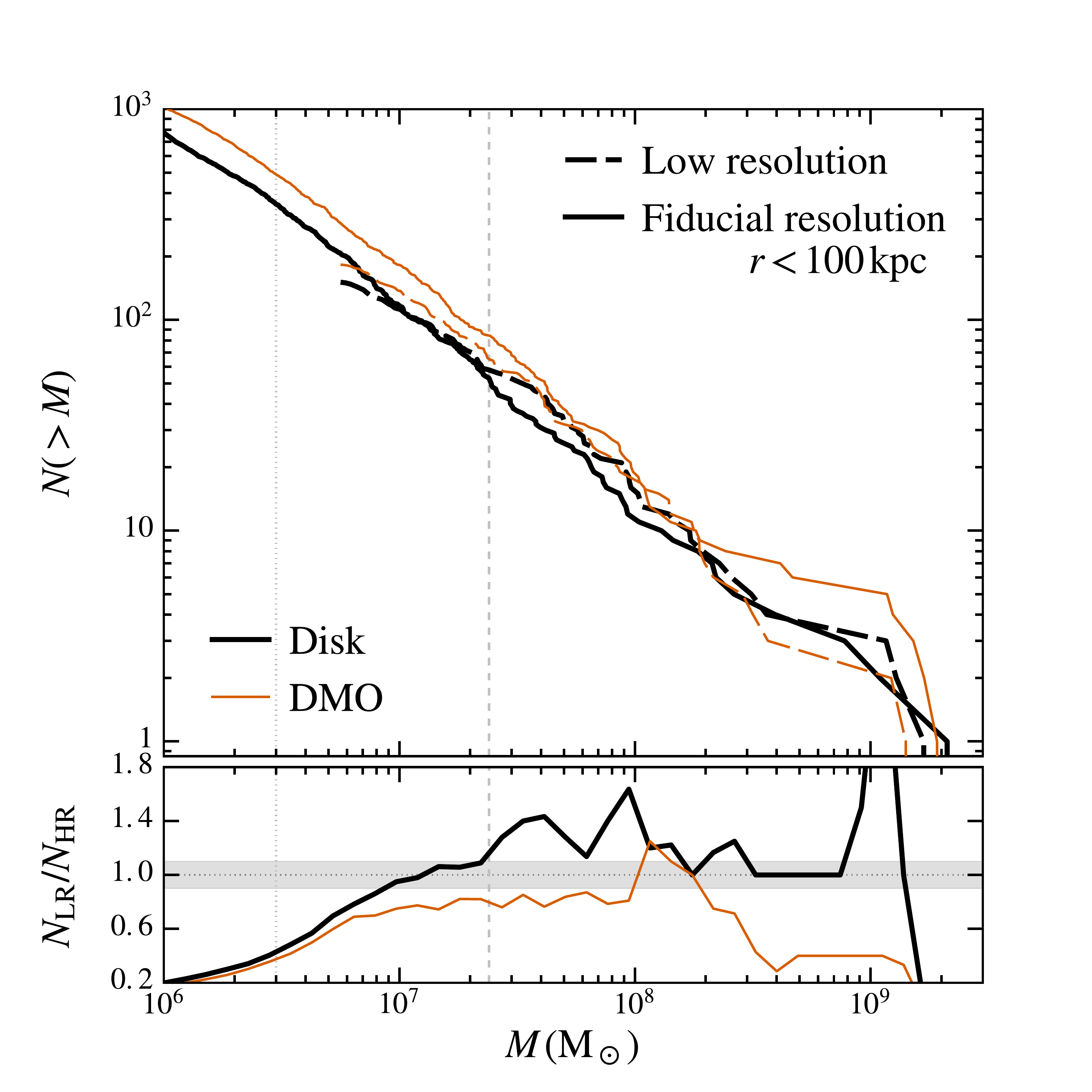}
    \includegraphics[width=\columnwidth]{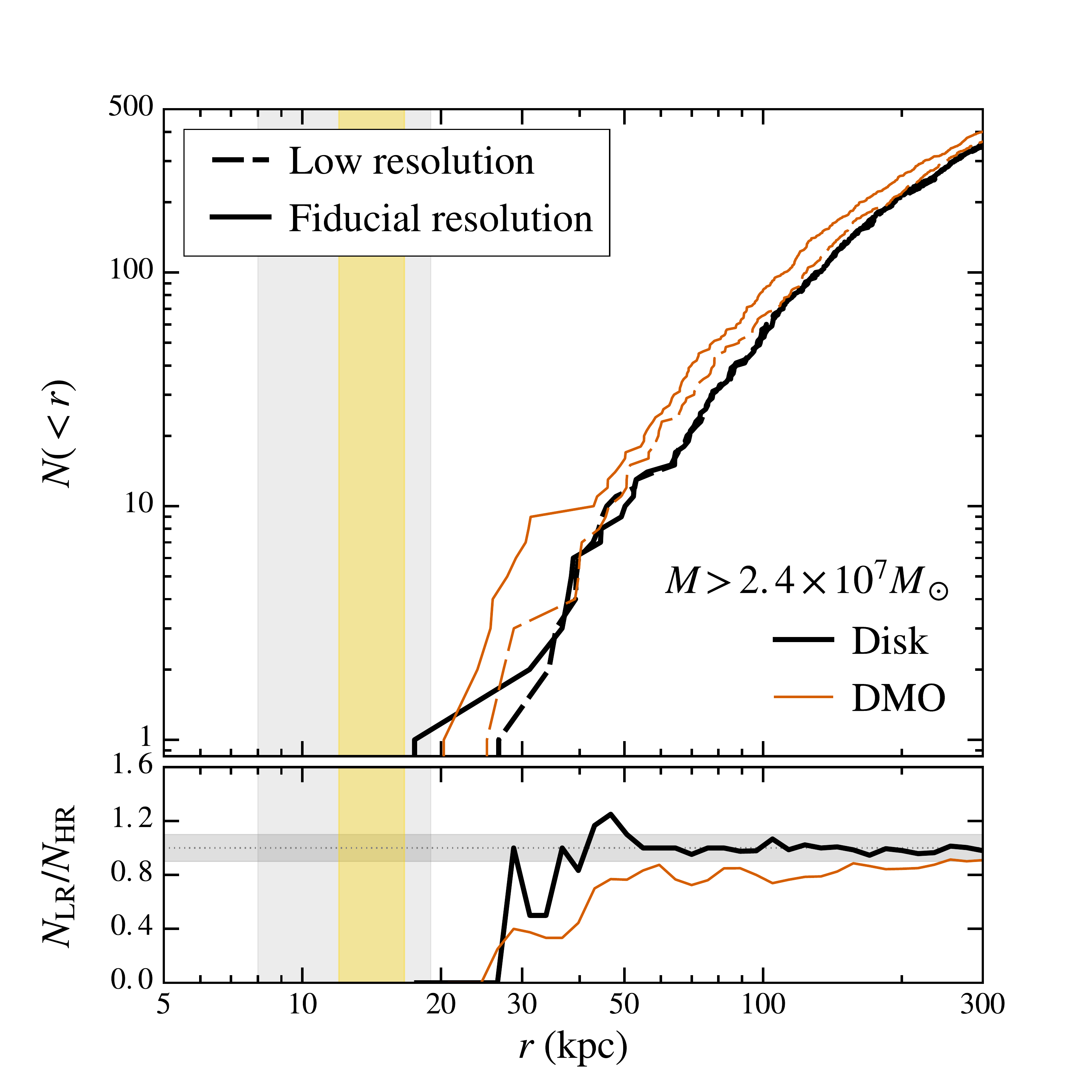}
\caption{
Comparing subhalo counts at our fiducial (high) resolution (solid lines) to those in a
low-resolution simulation with $8 \times$ fewer particles (dashed lines). The black
curves show the dark matter only (DMO) simulations while the orange curves show the
simulations with an embedded disk. The \emph{left} panel shows the cumulative number
of subhalos above a given subhalo bound mass, $M$, within $100~\kpc$.  We correct $M$
for the baryon fraction.  The dotted vertical line at $3\times10^6\msun$ shows the
resolution cut used in Figures~\ref{fig:massfuncs} -- \ref{fig:velocities}; the dashed
vertical line shows an equivalent number of particles in the low-resolution simulation,
$M = 2.4\times10^7\msun$. As demonstrated by the lower panel, which plots the ratio of
subhalo counts in the low- and high-resolution simulations, the counts are converged to
within $\sim15\%$ at that particle count. The shaded band indicates $\pm10\%$.  The
\emph{right} panel demonstrates that counts of subhalos above this mass are also well
converged as a function of radius:  the embedded disk simulations agree to within a few
percent at nearly all radii, and the relatively flat offset in the DMO run is consistent
with the underprediction discussed in the text.  The most significant deviations occur at
$r \lesssim 40 \kpc$ or $M > 3 \times 10^8 \Msun$, where the number of subhalos is small
($< 10$) and therefore subject to significant scatter.
}
\label{fig:resolution}
\end{figure*}

The left panel of Figure~\ref{fig:resolution} shows cumulative counts
as a function of bound mass assigned by \texttt{AHF}, $M$, within $100~\kpc$ of
\mi, corrected for $\fb$.  The black lines present counts in simulations with
the fiducial disk parameters, while the brown lines plot the DMO counterparts;
fiducial resolution simulations are plotted in solid, and those at lower
resolution are dashed.  The lower sub-panel shows the ratio between the
low and high resolution versions of the run with and without the disk.  The
dotted vertical line represents the resolution cut of $3\times10^6\msun\sim85$
particles adopted in the main text based on inspection of the differential
mass functions.  The dashed vertical line indicates a factor of $8$ larger
mass, corresponding to an identical number of particles in the lower
resolution simulation ($M = 2.4\times10^7\,\msun$, after multiplying by $1-\fb$).
Counts in the disk simulations agree well at that mass (though there
are fluctuations at higher mass), and counts in the DMO simulations only
differ by $\sim15\%$.  Convergence is generally even better on larger scales:
counts agree to within $10\%$ at nearly all $M > 2.4\times10^7\msun$.  The
low-resolution DMO simulation does, however, underpredict the subhalo count within
$300~\kpc$ by $\sim10\%$, and only agrees for $\sim300$ particles; at that particle
count, however, the low resolution disk simulation \emph{overpredicts} the subhalo
count by $\sim10\%$.

The right panel of Figure~\ref{fig:resolution} shows the radial distribution
of subhalos with $M > 2.4\times10^7\msun$ in the same four simulations.  Counts
in the disk runs agree to within a few percent at nearly all radii, and the DMO
simulations are within $\sim10$--$15\%$ of one another.  The offset in the latter
is roughly constant with radius, indicating that the deficiency discussed above
is relatively independent of distance. The most significant deviations occur at
$r \lesssim 40 \kpc$ or $M > 3 \times 10^8 \Msun$, where the number of subhalos is
small ($< 10$) and therefore subject to significant scatter, depending the subhalos'
orbital phases.  In general, the differences between the resolution levels are small
compared to the differences between the DMO and disk simulations.

\vspace{-1em}
\bibliographystyle{mnras}
\bibliography{disk_potential}

\bsp	
\label{lastpage}
\end{document}